\documentclass[aps,prd,superscriptaddress,groupedaddress,preprintnumbers,floatfix,twocolumn,nofootinbib]{revtex4}  

\usepackage{graphicx,comment}  
\usepackage{dcolumn}   
\usepackage{bm}        
\usepackage{amssymb}   

\usepackage{subfigure}
\usepackage{multirow}
\usepackage{placeins}
\usepackage{epstopdf}
\usepackage{mathtools}
\usepackage{color}

\usepackage{amsfonts}    
\usepackage{amsmath}
\usepackage{dcolumn}

\newcolumntype{d}[1]{D{.}{.}{1.8}}

\newcommand{\Ep}{E'^{*}}
\newcommand{\DD}{\Delta}

\begin{document}

\preprint{
\vbox{
\hbox{MIT-CTP/4894}
}}

\title{Off-forward gluonic structure of vector mesons
}

\author{W.~Detmold}\affiliation{Center for Theoretical Physics, Massachusetts Institute of Technology, Cambridge, MA 02139, U.S.A.}
\author{D.~Pefkou}\affiliation{Center for Theoretical Physics, Massachusetts Institute of Technology, Cambridge, MA 02139, U.S.A.}
\author{P.~E.~Shanahan}\affiliation{Center for Theoretical Physics, Massachusetts Institute of Technology, Cambridge, MA 02139, U.S.A.}
	
\begin{abstract}
The spin-independent and transversity generalised form factors (GFFs) of the $\phi$ meson are studied using lattice QCD calculations with light quark masses corresponding to a pion mass $m_\pi\sim450(5)$~MeV. One transversity and three spin-independent GFFs related to the lowest moments of leading-twist spin-independent and transversity gluon distributions are obtained at six non-zero values of the momentum transfer up to 1.2 GeV$^2$. These quantities are compared with the analogous spin-independent quark GFFs and the electromagnetic form factors determined on the same lattice ensemble. The results show quantitative distinction between the spatial distribution of transversely polarised gluons, unpolarised gluons, and quarks, and point the way towards further investigations of the gluon structure of nucleons and nuclei.
\end{abstract}

\maketitle

\section{Introduction}

Understanding the quark and gluon structure of hadrons and nuclei is a fundamental and compelling goal of nuclear physics. Over the last 60 years there has been extraordinary progress in both experimental measurement and theoretical understanding of the distributions of quarks inside hadrons and nuclei. 
There are now precise pictures of both the electromagnetic form factors and the quark distributions of nucleons and nuclei.
Determining the gluonic structure of these objects to a similar level, however, requires a new generation of experiments with higher luminosity and better detectors. In particular, an electron-ion collider (EIC) designed to fulfil these needs is currently in the planning phase~\cite{Accardi:2012qut}. Such a machine will provide access to a host of information about the gluonic structure of hadrons and nuclei, including transverse-momentum dependent distributions (TMDs) and gluon generalised parton distributions (GPDs). While many aspects of gluonic structure can be investigated, generalised transversity gluon distributions are of particular interest since they are purely gluonic; they do not mix with quark distributions at leading twist. In addition, the forward limit of these quantities are the double helicity flip parton distributions introduced by Jaffe and Manohar~\cite{Jaffe:1989xy}, which provide a clean signature of non-nucleonic degrees of freedom in nuclei of spin $\ge$ 1. Away from the forward limit, these distributions exist for targets of any spin~\cite{Hoodbhoy:1998vm,Belitsky:2000jk}.
 
 In this work the generalised gluon distributions (gluon GPDs) of the spin-1 $\phi$ meson are investigated. In particular, the generalised form factors (GFFs) corresponding to the first Mellin moments of the unpolarised and transversity gluon GPDs are determined for the first time using lattice QCD (LQCD) calculations, albeit at unphysical quark masses. The forward limits of these form factors correspond to the gluon momentum fraction and a transverse momentum asymmetry, respectively. Since the transversity gluon GPD is non-zero in the forward limit  only in targets of spin$\ge$1, the $\phi$ meson, which is the simplest spin-1 system, is chosen for this exploratory study. 
 
Because of the large number of both transversity and unpolarised GFFs that contribute to the first Mellin moments of the GPDs away from the forward limit, only a subset can be cleanly determined within the technical limitations of the LQCD calculation presented here. Nevertheless, three of the seven unpolarised gluon GFFs, and one of the eight transversity GFFs, are extracted for six non-zero momentum transfers in the range $0< |\Delta^2|< 1.2$~GeV$^2$. 
The unpolarised gluon GFFs are compared with the analogous quark GFFs, to which they have a one-to-one correspondence. While there are clear quantitative differences between the unpolarised gluon and quark distributions, and also between these and the gluon transversity distribution, interpreting these differences is a challenging problem. Resolving a full three-dimensional picture of the gluon structure of the $\phi$ meson will require more precise calculations that extend to more than the lowest moment of the GPDs.

This work represents the first probe of detailed aspects of the gluonic structure of a hadron using LQCD. A number of technical aspects of the study are novel, and the improvements presented here set the stage for future studies that will map out the full gluonic structure of the $\phi$ meson and other hadrons. Applied to nucleons and nuclei, these techniques will set QCD benchmarks for an EIC.

\section{Gluon GFFs for spin-1 particles}

GPDs encode the three-dimensional quark and gluon structure of hadrons and nuclei. They encompass the information carried by the parton distribution functions and the elastic electromagnetic form factors, describing the distribution of partons both in the transverse plane and in the longitudinal direction~\cite{Diehl:2003ny}. 
Through the operator product expansion, the towers of Bjorken-$x$ (Mellin) moments of the GPDs are related to matrix elements of towers of local twist-two operators. These matrix elements, in turn, are parametrised in terms of the GFFs which are the focus of this work.

There are three towers of moments of twist-2 gluon GPDs, encoding the spin-independent, spin-dependent and transversity distributions. These moments are related to matrix elements of the operators
\begin{align}\label{eq:SIop}
\overline{\mathcal{O}}_{\mu\nu\mu_1\ldots\mu_n} & =S\left[ G_{\mu\alpha}i\overleftrightarrow{D}_{\mu_1}\ldots i\overleftrightarrow{D}_{\mu_n}G_{\nu}^{\,\,\,\alpha}\right],\\\label{eq:SDop}
\tilde{\mathcal{O}}_{\mu\nu\mu_1\ldots\mu_n} & =S\left[ \tilde{G}_{\mu\alpha}i\overleftrightarrow{D}_{\mu_1}\ldots i\overleftrightarrow{D}_{\mu_n}G_{\nu}^{\,\,\,\alpha}\right],\\\label{eq:TTop}
\mathcal{O}_{\mu\nu\mu_1\ldots\mu_n} & =S\left[ G_{\mu\mu_1}i\overleftrightarrow{D}_{\mu_3}\ldots i\overleftrightarrow{D}_{\mu_n}G_{\nu\mu_2}\right],
\end{align}
respectively, where the gluon field strength tensor is $G_{\mu\nu}$, the dual field-strength tensor is $\tilde{G}_{\mu\nu}=\frac{1}{2}\epsilon_{\mu\nu\alpha\beta}G^{\alpha\beta}$, and $\overleftrightarrow{D}=\frac{1}{2}\left(\overrightarrow{D}-\overleftarrow{D}\right)$. `S' denotes symmetrisation and trace-subtraction in all free indices for Eqs.~\eqref{eq:SIop} and \eqref{eq:SDop}, and symmetrisation in the $\mu_i$ and and trace-subtraction in all indices for Eq.~\eqref{eq:TTop}. 
The matrix elements of these operators in spin-1 states, at lowest $n$, are the focus of this work.

The off-forward matrix elements of the twist-2 operators defined above are described by GFFs. 
The spin-dependent gluon GFFs, which vanish at lowest-$n$ through operator symmetries, are not considered numerically in this work but are enumerated in Appendix~\ref{app:GFFs}. 
With the polarisation vectors of massive spin-1 particles defined in Minkowski space as
\begin{equation}\label{eq:MinkPol}
E^\mu(\vec{p},\lambda) = \left( \frac{\vec{p}\cdot\vec{e}_\lambda}{m}, \vec{e}_\lambda + \frac{\vec{p}\cdot\vec{e}_\lambda}{m(m+\mathcal{E})}\vec{p}\right),
\end{equation}
where $\lambda=\{+,-,0\}$, $m$ and $\mathcal{E}=\sqrt{|\vec{p}|^2+m^2}$ are the rest mass and energy of the state, and
\begin{align}
\vec{e}_\pm & = \mp \frac{1}{\sqrt{2}}(0,1,\pm i), \\
\vec{e}_0 & = (1,0,0),
\end{align}
\begin{widetext}
\noindent
the spin-independent gluon GFFs are defined\footnote{This choice of basis is slightly different from that in Ref.~\cite{Taneja:2011sy}, where the decomposition also includes a trace term. Note also that not all GFFs are independent for all $n$.}~\cite{Taneja:2011sy} through 
\begin{align}
\nonumber
\left\langle p' E' \left|S \left[ G_{\mu\alpha} i\overleftrightarrow{D}_{\mu_1}\right. \right. \right.& \left.\left. \left. \hspace{-3mm}\ldots  i\overleftrightarrow{D}_{\mu_n}G_{\nu}^{\,\,\,\alpha} \right]\right| p E \right\rangle  \\
\nonumber
= 
\sum_{\substack{m \text{ even}\\m=0}}^n\left\{ \vphantom{\frac{B^{(n)}_{7,m}(\Delta^2)}{M^2}}\right.&\left.B^{(n+2)}_{1,m}(\Delta^2)M^2S\left[ E_\mu \Ep_\nu\DD_{\mu_1}\ldots\DD_{\mu_{m}}P_{\mu_{m+1}}\ldots P_{\mu_n}\right] \right.\\
\nonumber
& + B^{(n+2)}_{2,m}(\Delta^2) S\left[ (E\cdot \Ep)P_\mu P_\nu \DD_{\mu_1}\ldots\DD_{\mu_{m}}P_{\mu_{m+1}}\ldots P_{\mu_n} \right]\\
\nonumber
&+ B^{(n+2)}_{3,m}(\Delta^2) S\left[ (E\cdot \Ep)\DD_\mu \DD_\nu \DD_{\mu_1}\ldots\DD_{\mu_{m}}P_{\mu_{m+1}}\ldots P_{\mu_n} \right] \\
\nonumber
&+ B^{(n+2)}_{4,m}(\Delta^2)S\left[\left((\Ep\cdot P)E_\mu  +(E\cdot P)\Ep_\mu  \right) P_\nu \DD_{\mu_1}\ldots\DD_{\mu_{m}}P_{\mu_{m+1}}\ldots P_{\mu_n}\right] \\
\nonumber
&+ B^{(n+2)}_{5,m}(\Delta^2)S\left[\left((\Ep\cdot P)E_\mu-(E\cdot P)\Ep_\mu \right) \DD_\nu \DD_{\mu_1}\ldots\DD_{\mu_{m}}P_{\mu_{m+1}}\ldots P_{\mu_n} \right] \\
\nonumber
&+ \frac{B^{(n+2)}_{6,m}(\Delta^2)}{M^2}S\left[(E\cdot P)(\Ep\cdot P)P_\mu P_\nu \DD_{\mu_1}\ldots\DD_{\mu_{m}}P_{\mu_{m+1}}\ldots P_{\mu_n} \right] \\
\nonumber
&+ \left.\frac{B^{(n+2)}_{7,m}(\Delta^2)}{M^2}S\left[(E\cdot P)(\Ep\cdot P)\DD_\mu \DD_\nu \DD_{\mu_1}\ldots\DD_{\mu_{m}}P_{\mu_{m+1}}\ldots P_{\mu_n} \right] \right\} \\ \label{eq:SIME}
&+ {\rm mod}(n,2)B_{4,n}^{(n+2)}((\Ep\cdot P) E_\mu + (E\cdot P)\Ep_\mu)\Delta_\nu\Delta_{\mu_1}\ldots \Delta_{\mu_n}
.
\end{align}
Here, $P=(p+p')/2$ is the average momentum and the momentum transfer is defined as $\Delta = p'-p$. `S' denotes symmetrisation and trace-subtraction in all free indices. Of these GFFs, only $B^{(n)}_{1,0}(\Delta^2)$ and $B^{(n)}_{2,0}(\Delta^2)$ contribute to forward-limit matrix elements. The renormalisation scheme and scale-dependence of the GFFs is suppressed here.
 
The transversity GFFs are defined through
\begin{align}
\nonumber
\left\langle p' E' \left|S \left[ G_{\mu\mu_1} \vphantom{\overset{\leftrightarrow}{D}}\right.\right.\right. & \left.\left.\left. \hspace{-3mm} i \overset{\leftrightarrow}{D}_{\mu_3} \ldots i\overset{\leftrightarrow}{D}_{\mu_n}G_{\nu\mu_2} \right] \right| p E \right\rangle  \\
\nonumber
= \, \sum_{\substack{m \text{ even}\\m=2}}^n \left\{ \vphantom{\sum_i^j}\right. 
&A^{(n)}_{1,m-2}(\Delta^2)\, S\left[(P_\mu E_{\mu_1}-E_\mu P_{\mu_1})(P_\nu \Ep_{\mu_2} - \Ep_\nu P_{\mu_2})\DD_{\mu_3}\ldots\DD_{\mu_{m}}P_{\mu_{m+1}}\ldots P_{\mu_n}\right]\\
\nonumber
&+A^{(n)}_{2,m-2} (\Delta^2)\,S\left[(\DD_\mu E_{\mu_1}-E_\mu \DD_{\mu_1})(\DD_\nu \Ep_{\mu_2} - \Ep_\nu \DD_{\mu_2})\DD_{\mu_3}\ldots\DD_{\mu_{m}}P_{\mu_{m+1}}\ldots P_{\mu_n} \right]\\
\nonumber
&+A^{(n)}_{3,m-2}(\Delta^2)\,S\left[\left((\DD_\mu E_{\mu_1}-E_\mu \DD_{\mu_1})(P_\nu \Ep_{\mu_2} - \Ep_\nu P_{\mu_2})-(\DD_\mu \Ep_{\mu_1}-\Ep_\mu \DD_{\mu_1})(P_\nu E_{\mu_2} - E_\nu P_{\mu_2})\right)\right.\\
\nonumber
& \left. \hphantom{+A^{(n)}_{3,m-2}(\Delta^2)\,SS} \times \DD_{\mu_3}\ldots\DD_{\mu_{m}}P_{\mu_{m+1}}\ldots P_{\mu_n} \right]\\
\nonumber
&+A^{(n)}_{4,m-2}(\Delta^2)\, S\left[(E_\mu \Ep_{\mu_1}-E_{\mu_1}\Ep_\mu)(P_\nu\DD_{\mu_2}-P_{\mu_2}\DD_{\nu})\DD_{\mu_3}\ldots\DD_{\mu_{m}}P_{\mu_{m+1}}\ldots P_{\mu_n}\right]\\
\nonumber
&+\frac{A_{5,m-2}^{(n)}(\Delta^2)}{M^2}S\left[\left((E\cdot P)(P_\mu \DD_{\mu_1}-\DD_\mu P_{\mu_1})(\DD_\nu \Ep_{\mu_2}-\Ep_{\nu}\DD_{\mu_2})\right.\right.\\
\nonumber
&\hspace{2.9cm}+\left.\left.(\Ep\cdot P)(P_\mu \DD_{\mu_1}-\DD_\mu P_{\mu_1})(\DD_\nu E_{\mu_2}-E_{\nu}\DD_{\mu_2})\right)\DD_{\mu_3}\ldots\DD_{\mu_{m}}P_{\mu_{m+1}}\ldots P_{\mu_n}\right]\\
\nonumber
&+\frac{A_{6,m-2}^{(n)}(\Delta^2)}{M^2}S\left[\left((E\cdot P)(P_\mu \DD_{\mu_1}-\DD_\mu P_{\mu_1})(P_\nu \Ep_{\mu_2}-\Ep_{\nu}P_{\mu_2})\right.\right. \\
\nonumber
&\hspace{2.9cm}-\left.(\Ep\cdot P)\left.(P_\mu \DD_{\mu_1}-\DD_\mu P_{\mu_1})(P_\nu E_{\mu_2}-E_{\nu}P_{\mu_2})\right)\DD_{\mu_3}\ldots\DD_{\mu_{m}}P_{\mu_{m+1}}\ldots P_{\mu_n}\right]\\
\nonumber
&+ \frac{A_{7,m-2}^{(n)}(\Delta^2)}{M^2}(E'^{*}\cdot E)S\left[(P_\mu \DD_{\mu_1}-\DD_\mu P_{\mu_1})(P_\nu \DD_{\mu_2}-\DD_\nu P_{\mu_2})\DD_{\mu_3}\ldots\DD_{\mu_{m}}P_{\mu_{m+1}}\ldots P_{\mu_n}\right]\\
\nonumber
&+\frac{A_{8,m-2}^{(n)}(\Delta^2)}{M^4}(E\cdot P)(\Ep\cdot P)S\left[(P_\mu \DD_{\mu_1}-\DD_\mu P_{\mu_1})(P_\nu \DD_{\mu_2} - \DD_\nu P_{\mu_2})\DD_{\mu_3}\ldots\DD_{\mu_{m}}P_{\mu_{m+1}}\ldots P_{\mu_n}\right]\left. \vphantom{\sum_i^j} \right\} \\
\nonumber
& \hspace*{-15mm}+ \sum_{\substack{m \text{ odd}\\m=3}}^n \left\{ \vphantom{\sum_i^j}\right. 
A_{9,m-2}^{(n)}(\Delta^2)
S\left[(P_\mu \DD_{\mu_1}-\DD_\mu P_{\mu_1})  (\Ep_\nu P_{\mu_2}E_{\mu_3} +E_\nu P_{\mu_2}\Ep_{\mu_3} - \Ep_{\mu_2} P_{\nu}E_{\mu_3} -E_{\mu_2} P_{\nu}\Ep_{\mu_3})\right. \\
\nonumber &  \hspace*{30mm}
\times\left.\DD_{\mu_4}\ldots\DD_{\mu_{m+1}}P_{\mu_{m+2}}\ldots P_{\mu_n}\right]
\\
\nonumber
& + \frac{A_{10,m-2}^{(n)}(\Delta^2) }{M^2}
\left[(P_\mu \DD_{\mu_1}-\DD_\mu P_{\mu_1})  (\Ep_\nu \Delta_{\mu_2}E\cdot P -E_\nu \Delta_{\mu_2}\Ep\cdot P -\Ep_{\mu_2} \Delta_{\nu}E\cdot P +E_{\mu_2} \Delta_{\nu}\Ep\cdot P)\right. \\
\nonumber &  \hspace*{30mm}
\times\left.
\DD_{\mu_3}\ldots\DD_{\mu_{m}}P_{\mu_{m+1}}\ldots P_{\mu_n}\right]
\\
\nonumber 
& 
+ \frac{A_{11,m-2}^{(n)}(\Delta^2) }{M^2}
\left[(P_\mu \DD_{\mu_1}-\DD_\mu P_{\mu_1})  (\Ep_\nu P_{\mu_2}E\cdot P +E_\nu P_{\mu_2}\Ep\cdot P
-\Ep_{\mu_2} P_{\nu}E\cdot P -E_{\mu_2} P_{\nu}\Ep\cdot P)\right. \\\label{eq:TTME} &  \hspace*{30mm}
\times\left.
\DD_{\mu_3}\ldots\DD_{\mu_{m}}P_{\mu_{m+1}}\ldots P_{\mu_n}\right]
\left. \vphantom{\sum_i^j} \right\}. 
\end{align}
where the polarisation vectors $E$ and the momenta $P$ and $\Delta$ are as defined above. Here, `S' denotes symmetrisation in the indices $\mu_i$ (the pairs \{$\mu$,$\mu_1$\} and \{$\nu$,$\mu_2$\} are antisymmetric), symmetrisation of $\mu$ and $\nu$, and trace-subtraction in all free indices. The construction of this decomposition and that of Eq.~\eqref{eq:SIME} follows from applying discrete symmetries and demanding the correct Lorentz structure. Only $A^{(n)}_{1,0}(\Delta^2)$ contributes to forward-limit gluon transversity matrix elements.
%
\end{widetext}

\section{Lattice QCD calculation}
\label{sec:analysis}

In this work, a single ensemble of isotropic gauge-field configurations is used to determine the matrix elements discussed above at lowest $n$. The solutions of the systems of equations generated by various choices of polarisations and momenta in Eqs.~\eqref{eq:SIME} and \eqref{eq:TTME} 
allow subsets of the GFFs to be extracted, as will be discussed in detail below.
Simulations are performed with $N_f=2+1$ flavours of dynamical quarks, with quark masses chosen such that\footnote{Throughout this work, the $\phi$ meson is assumed to have a flavour content that is purely $\overline{s}s$ and annihilation contributions are ignored in two and three-point correlation functions. Such terms are suppressed by the Zweig rule.} $m_\pi\sim450(5)$~MeV and $m_\phi\sim 1040(3)$~MeV. A clover-improved quark action~\cite{Sheikholeslami:1985ij} and L\"uscher-Weisz gauge action~\cite{LuscherWeisz} are used, with the clover coefficient set equal to its tree-level tadpole-improved value. The lattices have dimensions $L^3\times T=24^3\times 64$, with lattice spacing $a=0.1167(16)$~fm~\cite{stefan}. Details of this ensemble are given in Table~\ref{tab:configs}~\cite{PhysRevD.92.114512}. 
\begin{table*}
	\begin{tabular}{ccccccccccccccc}\toprule
		$L/a$ & $T/a$ & $\beta$ & $am_l$ & $am_s$ & $a$~(fm) & $L$~(fm) & $T$~(fm) & $m_\pi$~(MeV) & $m_K$~(MeV) & $m_\phi$~(MeV) & $m_\pi L$ & $m_\pi T$ & $N_\textrm{cfg}$ & $N_\textrm{meas}$ \\ \hline
		24 & 64 & 6.1 & -0.2800 & -0.2450 & 0.1167(16) & 2.801(29) & 7.469(77) & 450(5) & 596(6) & 1040(3) & 6.390 & 17.04 & 1042 & $10^5$ \\\toprule
	\end{tabular}
	\caption{\label{tab:configs}
		LQCD simulation details.
		The gauge configurations have dimensions $L^3\times T$, lattice spacing $a$, and bare quark masses $a m_q$ (in lattice units). A total of $N_\textrm{meas}$ light-quark sources were used to perform measurements across $N_\textrm{cfg}$ configurations.}
\end{table*}

\subsection{Lattice operator construction}
\label{sec:latticeops}

The lowest-$n$ operators of the towers given in Eqs.~\eqref{eq:SIop} and \eqref{eq:TTop} are considered here.
Symmetrised and trace-subtracted, the Minkowski-space gluonic transversity operator for $n=2$ (Eq.~\eqref{eq:TTop}) does not mix with quark-bilinear operators of the same or lower dimension under renormalisation. The spin-independent gluonic operator with $n=0$ (Eq.~\eqref{eq:SIop}), however, mixes with the flavour singlet quark operator $\sum_{f=\{u,d,s\}}S\left[\overline{\psi}_f\gamma_\mu \overleftrightarrow{D_\nu}\psi_f\right]$, as discussed in more detail below.
Moreover, the discrete symmetries of a hypercubic lattice reduce the Lorentz group to the hypercubic group H(4), creating the possibility of further mixing.
Lattice operators with the appropriate continuum behaviour that do not have additional mixing with lower or same-dimensional operators were constructed, for the cases considered here, in Refs.~\cite{Gockeler:1996mu} and \cite{Detmold:2016gpy}. 

For the gluon transversity operator in Eq.~\eqref{eq:TTop}, operators in two irreducible representations of H(4) that do not mix with operators of same or lower dimension are investigated. Lattice operators which define bases of these representations are given explicitly in Appendix~\ref{app:latticeOps} along with their Minkowski-space analogues. These operators are constructed using the clover definition of the gluon field strength tensor, with gradient flow~\cite{Luscher:2010iy} applied to the links in the lattice gluon operators. The results shown use operators flowed to a total time of 1 in lattice units using a step size of 0.01.

As discussed in Ref.~\cite{Detmold:2016gpy}, the transversity lattice operators are related to continuum Euclidean-space operators through a finite multiplicative renormalisation factor:
\begin{equation}
\mathcal{O}^{(E)}_{l,m,n}=Z_{l,m} O^{\text{latt}}_{l,m,n},
\end{equation}
where the subscript $(l,m,n)$ denotes the $n$th vector from the $m$th representation of a lattice operator (where such operators are identified by the subscript $l$), and $Z_{l,m} = 1+\mathcal{O}(\alpha_s)$. 
In this work the renormalisation factors are not computed, but it is expected based on studies of similar gluonic operators~\cite{Alexandrou:2016ekb} that they are $\mathcal{O}(1)$.

Lattice operators from two representations of H(4) are considered for the spin-independent gluon operator defined in Eq.~\eqref{eq:SIop}, with explicit definitions given in Appendix~\ref{app:latticeOps}. 
As noted above, these operators mix with the quark operator $\sum_fS\left[\overline{\psi}_f\gamma_{\mu} \overleftrightarrow{D_{\nu}}\psi_f\right]$.
With the lattice operators corresponding to this quark bilinear denoted by $\overline{\mathcal{Q}}^{\text{latt}}_{l,m,n}$ with subscripts defined as above (where operators transforming irreducibly under H(4) are constructed in the same way as those for the corresponding gluonic operator), this mixing under renormalisation can be expressed as
\begin{equation}
\overline{\mathcal{O}}_{l,m,n}^{(E)} = Z^{gg}_{l,m} \overline{\mathcal{O}}^{\text{latt}}_{l,m,n}+Z^{gq}_{l,m} \overline{\mathcal{Q}}^{\text{latt}}_{l,m,n}.
\end{equation}
In Ref.~\cite{Alexandrou:2016ekb} it is shown numerically that this mixing, i.e., the magnitude of $Z^{qg}_{l,m}$, is at the few-percent level for a similar action to the one used here, and that the renormalisation $Z^{gg}_{l,m}$ is approximately unity, with several levels of stout smearing used on the operator. These small mixing effects are neglected in the present calculation.

\subsection{Determination of matrix elements}
\label{sec:MEs}

Matrix elements of the operators discussed in the previous section in the $\phi$ meson can be extracted from ratios of two and three-point correlation functions. With $\eta_j(\vec{p},t)$ denoting the vector of $\phi$ interpolating operators and where $\epsilon_j$ are Euclidean polarisation vectors related to the Minkowski expression in Eq.~\eqref{eq:MinkPol} by $\epsilon_j(\vec{p},\lambda) = E^j(\vec{p},\lambda)$, such that\footnote{Note that cubic symmetry guarantees the polarisation-independence of $Z_\phi(\vec{p})$.}
\begin{equation}
\left\langle 0 |\eta_j(\vec{p})|\vec{p},\lambda\right\rangle = Z_\phi(\vec{p})\, \epsilon_j(\vec{p},\lambda),
\end{equation}
where $\vec{p}$ is the momentum of a state and $\lambda$ labels its polarisation,
the two-point function can be expressed as
\begin{align}\nonumber
C_{jk}^{2\text{pt}}(\vec{p},t)&=\left\langle \eta_k(\vec{p},t)\eta^\dagger_j(\vec{p},0)\right\rangle\\
&=|Z_\phi(\vec{p})|^2 \left( e^{-\mathcal{E}t} + e^{-\mathcal{E}(T-t)}\right)\sum_{\lambda} \epsilon_k(\vec{p},\lambda)\epsilon^{*}_j(\vec{p},\lambda).
\end{align}
Contributions from excited states (which are exponentially suppressed) are omitted from this expression. In analysis, care is taken to restrict to time ranges where such contamination is negligible. 

Three-point correlation functions are constructed by taking the correlated product, configuration-by-configuration and source-location--by--source-location, of these two-point functions\footnote{For three-point functions with off-diagonal polarisations in the helicity basis, the two-point functions used are zero when an ensemble average is taken, but signals emerge through their correlation with the gluonic operators. The gluon transversity operators considered here are themselves zero on ensemble average.} with the gluonic operators calculated as described in the previous section. 
While the only case for which the vacuum expectation value $\left\langle \eta_k(\vec{p}\,',t) \, \eta^\dagger_j(\vec{p},0)\right\rangle\left\langle \vphantom{\eta^\dagger_j} \mathcal{O}(\vec{p}\,'-\vec{p},\tau)\right\rangle$ is non-zero is where $\vec{p}=\vec{p}\,'$ and $\mathcal{O}$ is a spin-independent gluon operator, a vacuum subtraction is performed for every operator and all momenta in this calculation. This correlated subtraction of zero improves the signal-to-noise ratio significantly.
Inserting complete sets of states, the subtracted three-point correlators can thus be expressed as
\begin{align}\nonumber
C_{jk}^{3\text{pt}}&(\vec{p},\vec{p}\,',t,\tau,\mathcal{O})\equiv\left\langle \eta_k(\vec{p},t) \,\mathcal{O}(\vec{p}\,'-\vec{p},\tau)\, \eta^\dagger_j(\vec{p}\,',0)\right\rangle\\\nonumber
& \hspace{2cm} - \left\langle \eta_k(\vec{p},t) \, \eta^\dagger_j(\vec{p}\,',0)\right\rangle\left\langle \vphantom{\eta^\dagger_j} \mathcal{O}(\vec{p}\,'-\vec{p},\tau)\right\rangle\\ \label{eq:3pt}
& = Z_\phi^\dagger(\vec{p})Z_\phi(\vec{p}\,') e^{-\mathcal{E}t} \sum_{\lambda \lambda'} \epsilon_k(\vec{p},\lambda)\epsilon^{*}_j(\vec{p}\,',\lambda')\langle \vec{p},\lambda|\mathcal{O}|\vec{p}\,',\lambda'\rangle
 \end{align}
for $0\ll\tau\ll t\ll T$ (where $T$ denotes the time extent of the lattice).  For the case $0\ll t\ll \tau\ll T$, $t$ is replaced by $(T-t)$ in the final line of the above expression and there is an additional multiplicative factor of $(-1)^{n_4}$, where $n_4$ is the number of temporal indices in the operator.

The two and three-point correlation functions were constructed from propagators computed using a bare quark mass $m = -0.2450$ and 5 iterations of gauge-invariant Gaussian smearing in the spatial directions at both source and sink, with interpolating operators of the form $\eta_j(x)=\overline{\psi}(x)\gamma_i\psi(x)$ in terms of smeared quark fields. On each of 1042 configurations, spaced by 10 trajectories, 96 source locations were used, and measurements were averaged over these source locations before a bootstrap analysis was performed to assess statistical uncertainties.

 The leading exponential time-dependence in Eq.~\eqref{eq:3pt}, as well as factors of $Z_\phi$, can be eliminated by forming the ratio:
\begin{align}\nonumber
R_{jk}(\vec{p},\vec{p}\,',t,\tau,\mathcal{O}) =&\\\label{eq:ratR} \frac{C^{3\text{pt}}_{jk}(\vec{p},\vec{p}\,',t,\tau,\mathcal{O})}{C_{kk}^{2\text{pt}}(\vec{p}\,',t)}&\sqrt{\frac{C_{jj}^{2\text{pt}}(\vec{p},t-\tau)C_{kk}^{2\text{pt}}(\vec{p}\,',t)C_{kk}^{2\text{pt}}(\vec{p}\,',\tau)}{C_{kk}^{2\text{pt}}(\vec{p}\,',t-\tau)C_{jj}^{2\text{pt}}(\vec{p},t)C_{jj}^{2\text{pt}}(\vec{p},\tau)}},
\end{align}
which is proportional, through factors of $m_\phi$ and momentum components, to the matrix elements of interest, $\langle \vec{p}\,',\lambda '|\mathcal{O}|\vec{p},\lambda\rangle$. For each lattice operator $\mathcal{O}$, this ratio was constructed for all diagonal and off-diagonal polarisation combinations $jk$, and for momenta up to $\vec{p}\,^2=4$ and $\vec{p}\,'^2=4$, taken in all combinations that give a resultant momentum transfer up to $|\vec{\Delta}|^2=(\vec{p}\,'-\vec{p}\,)^2 =6$. Ratios for the cases with $\tau \ll t$ and $t \ll \tau$ are averaged, with the appropriate signs included as discussed following Eq.~\eqref{eq:3pt}.

\subsection{Extraction of GFFs}
\label{subsec:extract}

 The Euclidean operators used here are given explicitly in Appendix~\ref{app:latticeOps} and discussed in Sec.~\ref{sec:latticeops}.
 Matrix elements of these operators, encoded in the ratios $R_{jk}(\vec{p},\vec{p}\,',t,\tau,\mathcal{O})$ described in the previous section, are matched to GFFs by applying Eqs.~\eqref{eq:SIME} and \eqref{eq:TTME} to the corresponding Minkowski-space operator expressions (also made explicit in Appendix~\ref{app:latticeOps}). For each basis of operators, at each value of the momentum transfer, this generates systems of equations for the GFFs, $B_{i,m}^{(2)}(\Delta^2)$ or $A_{i,m}^{(2)}(\Delta^2)$ depending on the operator, where each equation corresponds to one choice of operator in the basis and one set of momentum and polarisation vectors. 
 In general, these systems are large, the relation between the GFFs and the matrix elements is not simply invertible, and the systems can not be solved for all GFFs simultaneously. This is discussed in detail below. 

The extraction of the GFFs proceeds in four of steps:
\begin{enumerate}
	\item Construct averages of the ratios $R_{jk}(\vec{p},\vec{p}\,',t,\tau,\mathcal{O})$ for equivalent choices of polarisation, momenta, and operators in a given basis;
	\item Fit constants to the averaged ratios in their plateau regions;
	\item Determine the GFFs which are dominant in the analysis for each operator;
	\item Solve the (possibly over or under-determined) system for the dominant GFFs at each $\Delta^2$;
\end{enumerate}
each of which will be described in detail.

For each operator under consideration, the numerical values of the ratios $R_{jk}(\vec{p},\vec{p}\,',t,\tau,\mathcal{O})$ are averaged, at the bootstrap level, over all choices of momentum and polarisation that give the same linear combination of GFFs up to a sign (by Eq.~\eqref{eq:SIME} or \eqref{eq:TTME} as appropriate). This procedure defines a reduced set of unique, but not linearly independent, linear equations for each operator and momentum transfer.

\begin{figure}
	\includegraphics[width=0.49\textwidth]{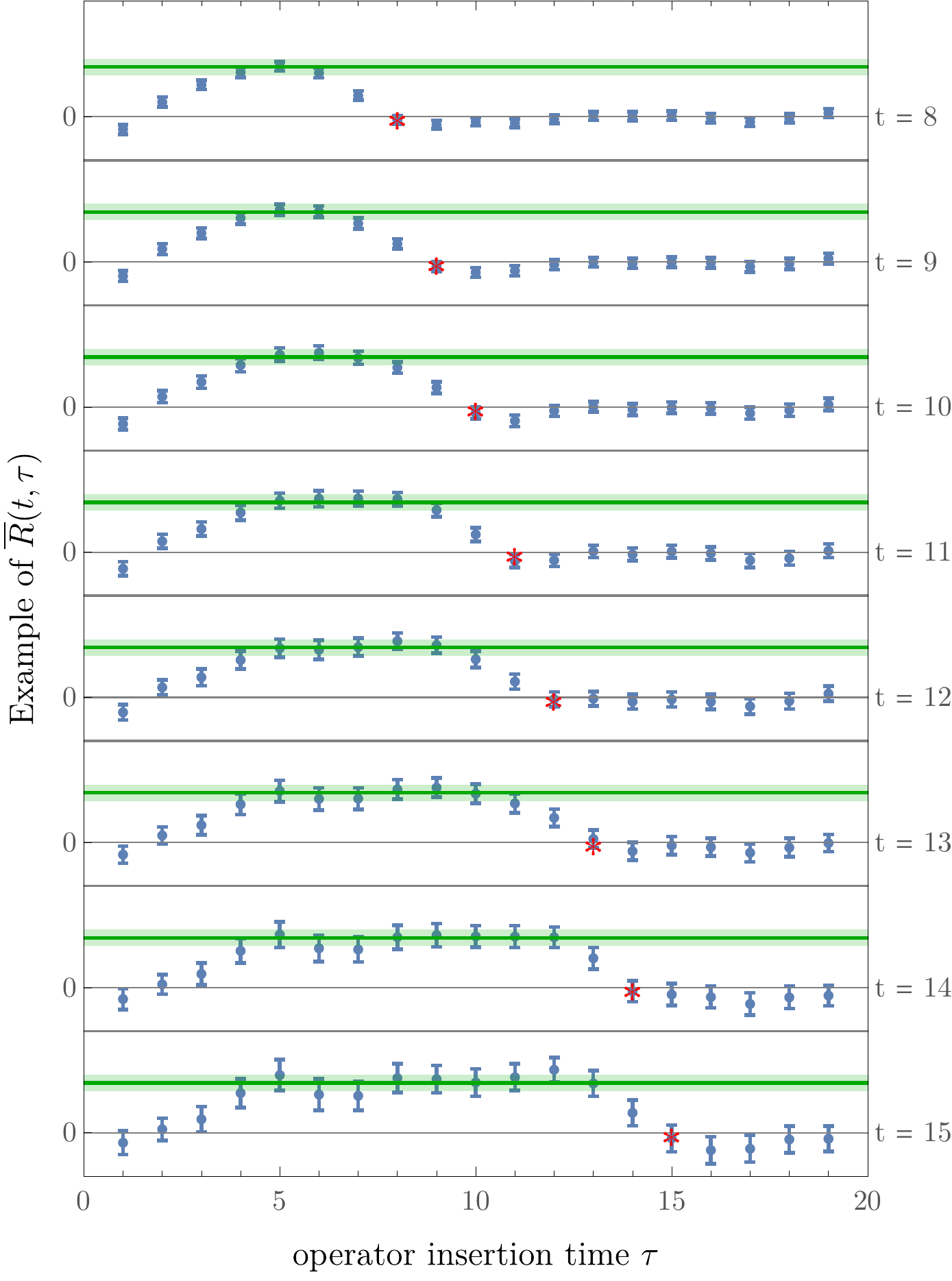}
	\caption{\label{fig:bumps}Example of a plateau fit to an averaged ratio $\overline{R}(t,\tau)$. Each section of the figure (separated by horizontal lines) shows $\overline{R}(t,\tau)$ plotted against operator insertion time $\tau$ at a fixed value of the sink time $t$, which is denoted by the red star on each plot. The result of a fit to the (two-dimensional) plateau region, determined as described in the text, is shown on each cross-section as a green horizontal band.}
\end{figure}

For each averaged ratio $\overline{R}(t,\tau)$, the maximal connected plateau region in $t$--$\tau$ space is determined, where this is defined as the region where the bootstrap-level differences between all pairs of adjacent points are consistent with zero. If this maximal plateau region for an averaged ratio consists of less than 10 $(t,\tau)$ pairs, that ratio is discarded from the analysis (typical fits include many more points in the plateau region).
Given this maximal plateau region, the variation in central values of fits to all sub-regions is taken as a measure of the fitting uncertainty, while the bootstrap-level fit of a constant over the maximal region gives the central value and statistical uncertainty of the fit. These uncertainties are propagated into the subsequent analysis as described below.
In this analysis, all $(t,\tau)$ combinations are available, so a comprehensive elimination of excited states can be achieved; this aspect of the calculation is better controlled than for studies of quark operators where each $\tau$ (or $t$) value requires additional computation and so typically only a few values can be used. Figure~\ref{fig:bumps} shows an example of such a plateau fit to an averaged ratio in the $t$--$\tau$ plane.

Because of the large number of GFFs that contribute to the off-forward matrix elements, not all can be determined from the LQCD calculations presented here. 
A complete extraction would require precise data from many different sets of initial and final momenta giving the same momentum transfer. This could be achieved either with new techniques allowing high-precision data to be obtained at large momenta~\cite{Chambers:2017tuf,Bali:2016lva,sergey}, or with very large lattices having allowed values of momentum transfer that are sufficiently closely spaced in physical units to allow binning. Given the sets of momenta available with good precision in these calculations, the linear systems generated by the matching of the LQCD results to the corresponding matrix elements in terms of GFFs do not contain enough independent equations to constrain all GFFs for some operators at some momentum transfers. 
 In other cases, the contributions from a number of the GFFs are suppressed by several orders of magnitude relative to others, again making the extraction of these quantities impossible with the current statistical precision. Moreover, for some bases of operators, symmetries relate the coefficients of two or more GFFs for every choice of momentum and polarisation, meaning that those GFFs can not be separated by any fit, regardless of the precision of the results or the number of momenta available.
 
 \begin{figure}
 	\subfigure[]{
 		\includegraphics[width=0.49\textwidth]{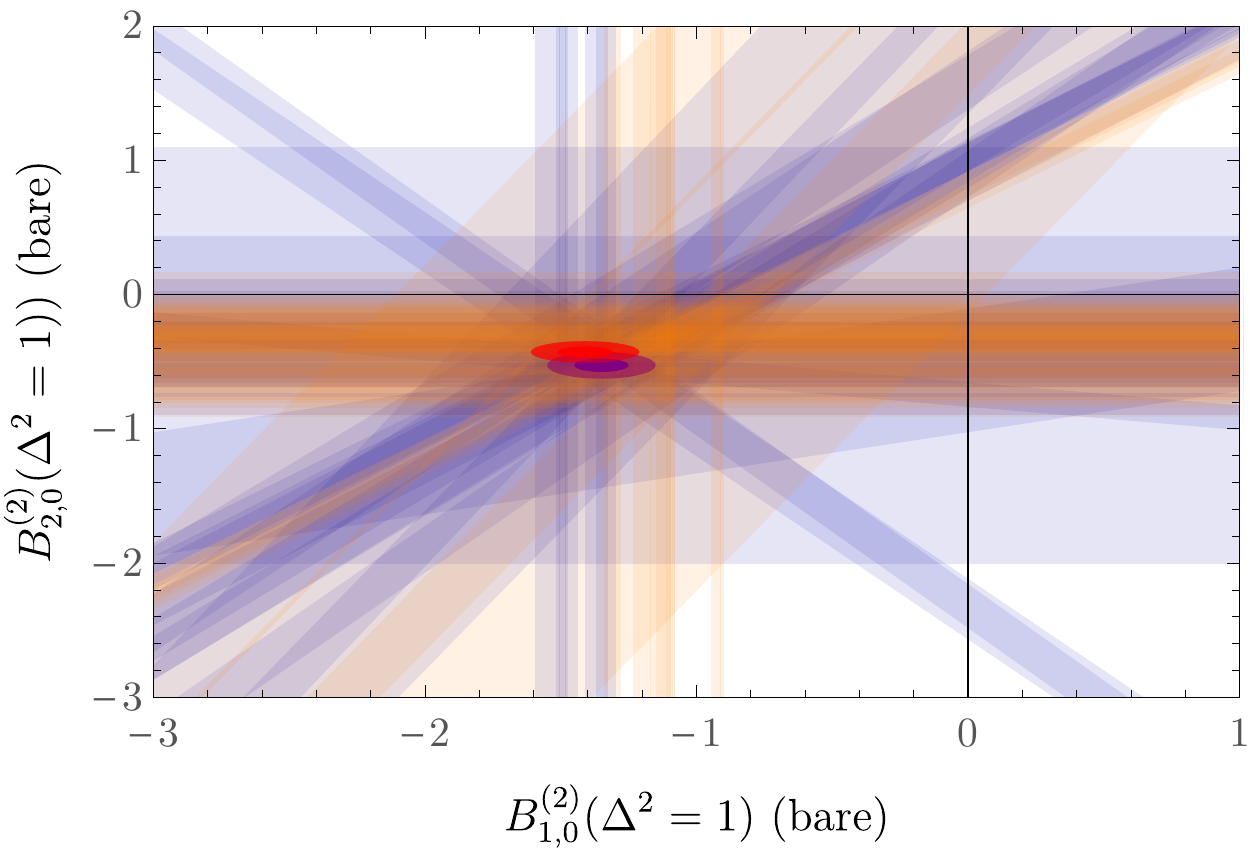}}
 	\subfigure[]{
 		\includegraphics[width=0.49\textwidth]{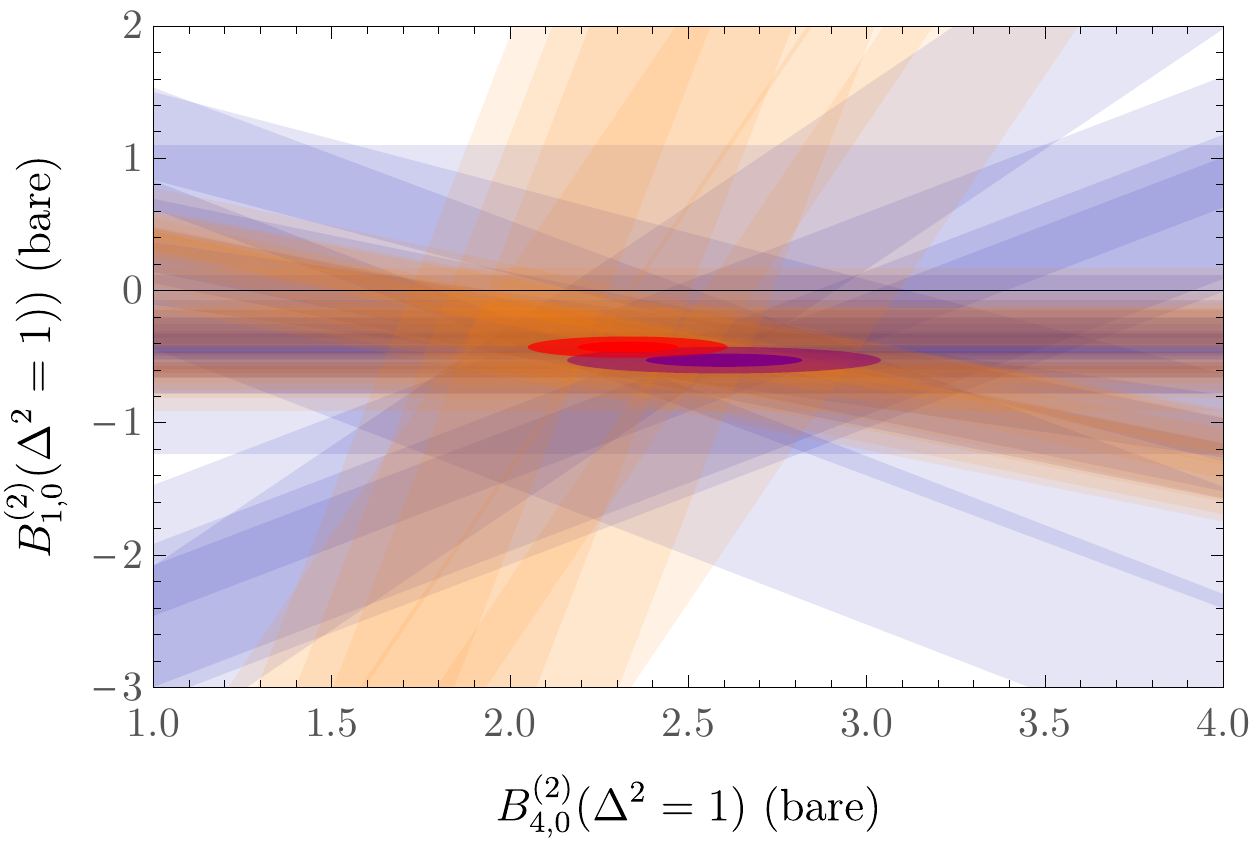}}
 	\subfigure[]{
 		\includegraphics[width=0.49\textwidth]{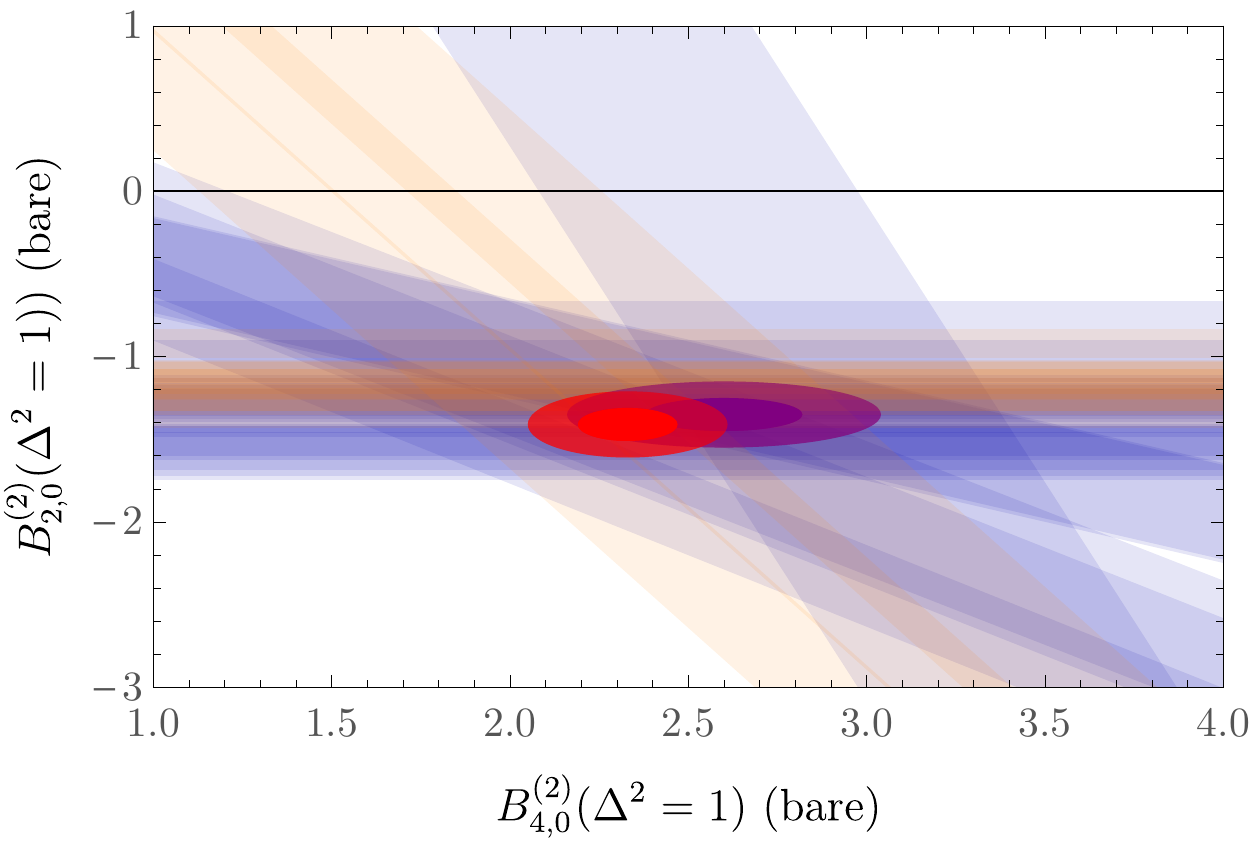}}
 	\caption{\label{fig:bands}Cross-sections of the multi-dimensional system of linear equations for the matrix elements of spin-independent gluonic operators in the $\tau_1^{(3)}$ (blue) and $\tau_3^{(6)}$ (orange) representations (see Appendix~\ref{app:latticeOps}), at the lowest non-zero momentum transfer. Each band shows the central value and uncertainty, defined as described in the text, corresponding to one linear equation in the system, with values of the GFFs $B^{(2)}_{i,0}(\Delta^2)$ with $i\ne \{1,2,4\}$ set to $0\pm10$, projected into the planes of the dominant GFFs. Bands with much larger uncertainties in this particular projection are omitted for clarity. In the infinite-statistics limit, all bands should intersect at a single point in the multi-dimensional space. The ellipses show the results obtained from the fits to the multidimensional systems as described in the text. }
 \end{figure}
 
 For each basis of operators, a subset of the GFFs can, however, be extracted. This set of dominant GFFs is found by inspection of the relative weights of each GFF in the system of linear equations to be solved. For example, for all gluon transversity basis operators considered, at all momentum transfers, the majority of equations in the linear systems have the coefficient of $A_{1,0}^{(2)}(\Delta^2)$ significantly larger than the coefficients of the other GFFs (see Appendix~\ref{app:lineqs}). For this reason, $A_{1,0}^{(2)}(\Delta^2)$ is considered to be the dominant GFF, and its extraction is the focus of this work. Similar arguments lead to three GFFs, namely $B^{(2)}_{1,0}(\Delta^2)$, $B^{(2)}_{2,0}(\Delta^2)$  and $B^{(2)}_{4,0}(\Delta^2)$,  being targeted in the spin-independent case. More GFFs are resolvable at some particular momenta, but the aim of this work is to obtain a subset of the GFFs that can be determined consistently at all momenta, as described below.
 \begin{figure*}
 	\centering
 	\subfigure[Basis 1: $\tau_1^{(2)}$]{\label{sf:TEGFFsa}
 		\includegraphics[width=0.48\textwidth]{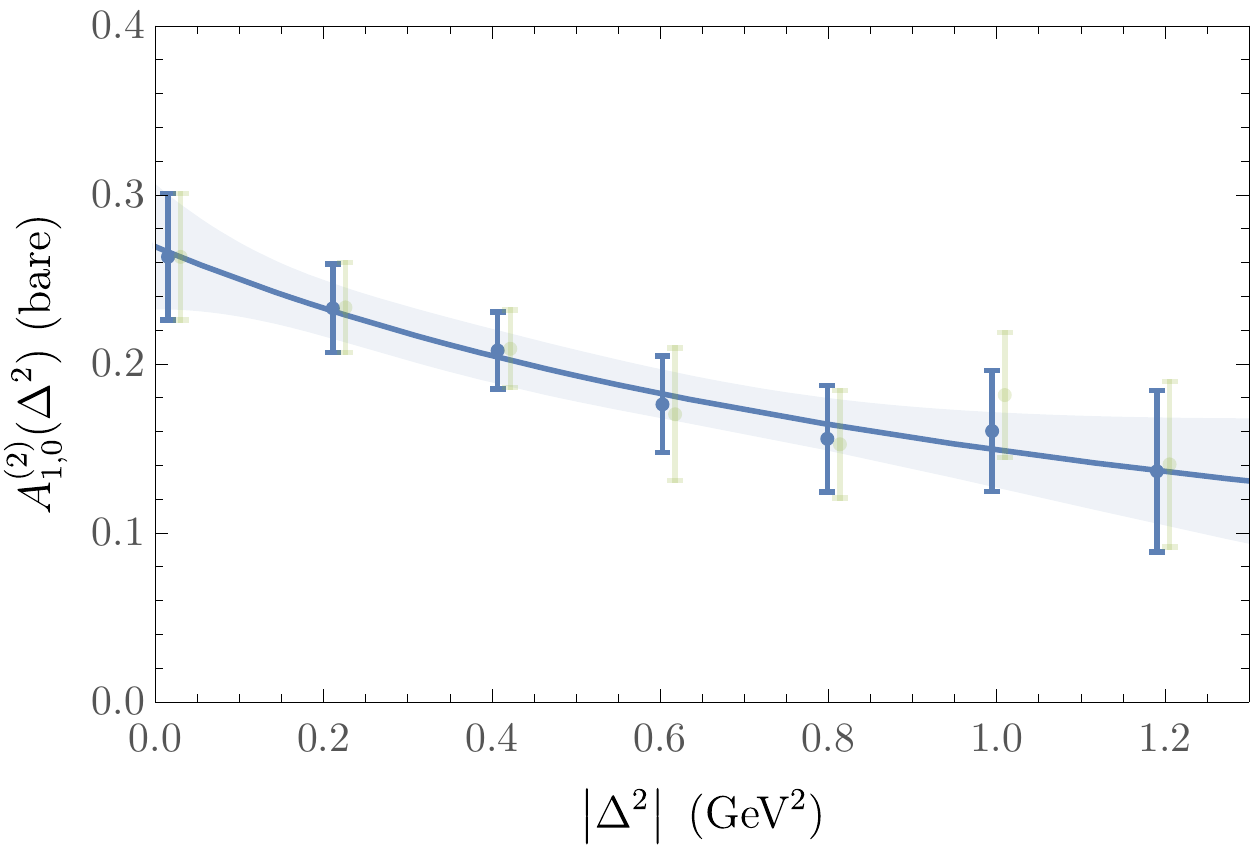}
 	}
 	\subfigure[Basis 2: $\tau_2^{(6)}$]{\label{sf:TEGFFsb}
 		\includegraphics[width=0.48\textwidth]{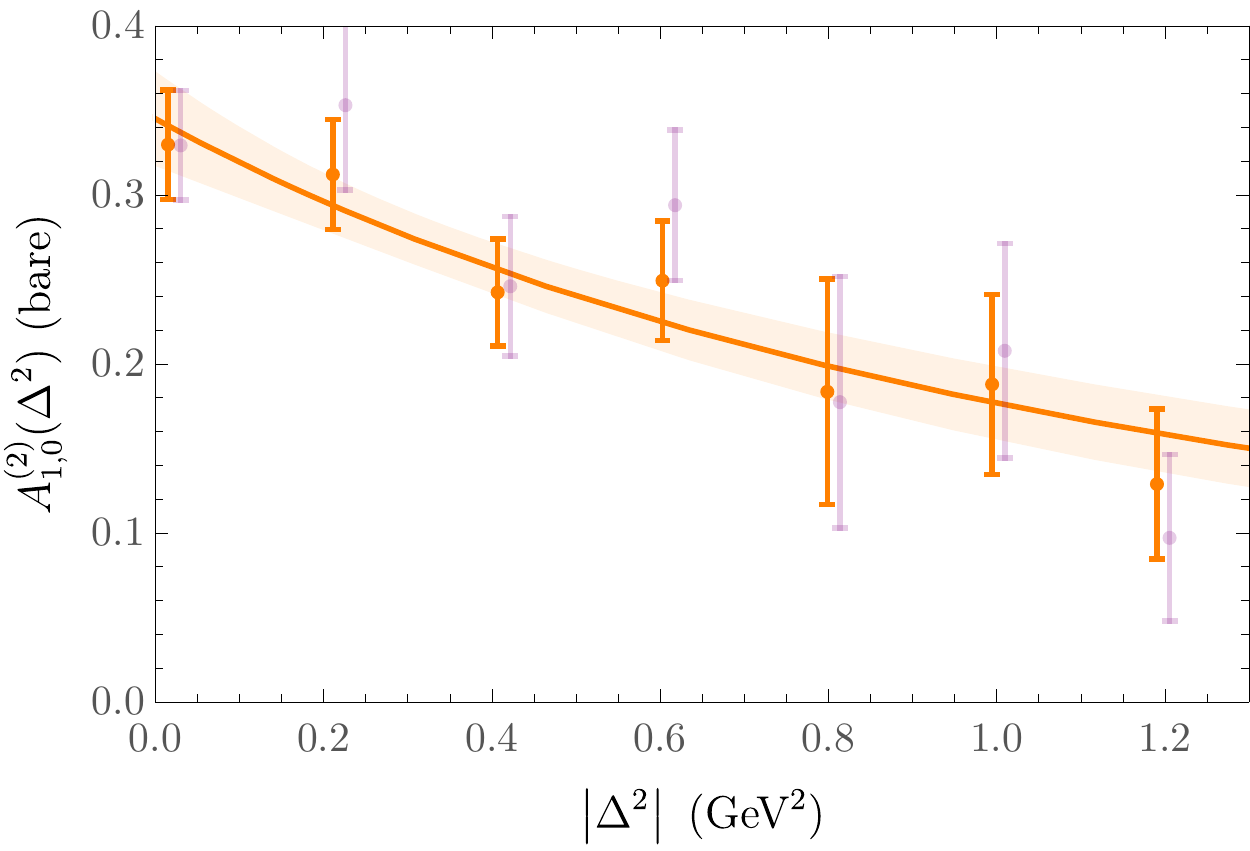}
 	}
 	\caption{\label{fig:TEGFFs}The gluon transversity GFF extracted from this analysis as described in Section~\ref{sec:analysis}. Subfigures~(a) and (b) show results obtained using lattice operators from the two different irreducible representations considered. The solid blue and orange points denote the results of the full analysis, while the faded green and purple points (offset on the horizontal axis for clarity) show the analysis repeated without the smoothness constraint discussed at the end of in Section~\ref{subsec:extract}. The bands are dipole fits to the results of the full analysis against $\left|\Delta^2\right|$, shown to guide the eye.}
 \end{figure*}

To achieve this, least-$\chi^2$ fits at the bootstrap level are performed to the systems of equations for each basis of operators, with fitting uncertainties assigned to each averaged ratio $\overline{R}$ as described above. Such fits are performed multiple times, fitting to every subset of the GFFs that includes those classed as dominant, with the GFFs which are not fit set to 0 and assigned an uncertainty of 10 in the $\chi^2$ fit\footnote{While there are no current bounds on the magnitudes of the GFFs to support this choice, an order of magnitude variation would be surprising. Moreover, as fits to all subsets of GFFs are included in the analysis, significant contributions outside of this bound from a ‘sub-dominant’ GFF would become apparent from inconsistencies between fits to different subsets. This is not observed.}. There are thus 16 sets of fits for the unpolarised operator and 128 sets of fits for the transversity operator. 
The variation in central values over the fits to different subsets with acceptable values of $\chi^2$ is included in quadrature as a second fitting systematic uncertainty on the final results, while the central values and statistical and plateau fitting uncertainties are taken from the bootstrap fits over the minimal set of dominant GFFs.

Precisely, for a fit to the subset of GFFs $f_{i\in S,j}$, where $S$ denotes the set of GFFs which are fit over and the subscript $j$ labels the discrete values of the momentum transfer $\Delta^2$, the first contribution to the $\chi^2$ function can be expressed as:
\begin{equation}
\chi^2_A(b,f_{i\in S,j})=\sum_{\overline{R}}\frac{\left(M_{\overline{R}}(f_{i\in S,j})|_{(f_{i\notin S,j}=0)}-\overline{R}(b)\right)^2}{\Delta\overline{R}^2+\hat{\Delta}(\overline{R})^2}.
\end{equation}
Here $\overline{R}(b)$ denotes the set of averaged plateau values extracted from the ratios discussed previously, for a given bootstrap, and $\Delta\overline{R}$ represents the statistical and fitting uncertainties on these quantities, determined as described earlier. The label $b$ indicates that this $\chi^2$ is formed for each bootstrap.
$M_{\overline{R}}$ represents the expectations, from Eq.~\eqref{eq:SIME} or \eqref{eq:TTME}, for the averaged ratio $\overline{R}$, in terms of the GFFs, where those GFFs not in the subset $S$ are set to zero. The quantity $\hat{\Delta}$ assigns an uncertainty of 10 to those GFFs which are not fit to: 
\begin{equation}
\hat{\Delta}(\overline{R})=\sqrt{\sum_{k\notin S} \left(M_{\overline{R}}(f_{i,j})\big|_{(f_{k,j}=10)}-
	M_{\overline{R}}(f_{i,j})\big|_{(f_{k,j}=0)}\right)^2}.
\end{equation}
Choosing an uncertainty of 100 gives entirely consistent results, albeit with larger uncertainties.

In the fits that are performed, an additional contribution is added to the $\chi^2$ function, representing a dipole function fit to each of the dominant GFFs as a whole. This addition has the effect of correlating the solutions of the systems of overdetermined equations at different values of $\Delta^2$, which results in GFFs that are somewhat more smoothly behaved and reduces the uncertainty on the $\left|\Delta^2\right|=\{2,3\}$ (in lattice units) points in particular. The particular $\Delta^2$ points that see improvement, fit alone, are less well constrained than others as there are less combinations of polarisations/momenta available.
Precisely, this additional contribution to the total $\chi^2$ can be expressed as 
\begin{equation}
\chi^2_B(b,f_{i,j},\mu_i,a_i)=\sum_{i\in S,j} \frac{\left(\frac{\mu_i}{(1+a_i\Delta^2)^2}-f_{i,j}  \right)^2}{\left(\frac{1}{2}\frac{\partial^2 \chi^2_A}{\partial f_{i,j}^2}\right)^2},
\end{equation}
where $\mu_i$ and $a_i$ are the parameters of the dipole fits to the GFF labelled by $i$.
Note that $\chi^2_A$ is a quadratic function of the GFFs $f_{i,j}$, so the denominator weights each term by how well the relevant GFF is determined. A correlation matrix is not used, since $\chi^2_A$ does not include cross-terms between different momenta.

Results are obtained by minimization of the total:
\begin{equation}
\chi^2_\text{tot}(b,f_{i,j},\mu_i,a_i) = \chi^2_A(b,f_{i,j})+\chi^2_B(b,f_{i,j},\mu_i,a_i),
\end{equation}
to determine the dominant GFFs at each momentum transfer. The mean and standard deviation over the GFFs determined on each bootstrap are used as the central value and first uncertainty on the quoted results.
In addition, the uncertainties on the GFFs determined by minimizing $\sum_b \chi^2_\text{tot}(b,f_{i,j},\mu_i,a_i)$ are included in quadrature. In most cases, where the bootstrap uncertainties accurately reflect the fitting uncertainties, this addition has little effect. In the few cases where sub-dominant form factors are significant at certain momenta (and setting them to be centered around 0 distorts the bootstrap fits), this addition inflates the fitting uncertainties considerably. 
Figure~\ref{fig:bands} illustrates one of the critical aspects of the fitting procedure, namely solving the linear system of constraints. The results in Figs.~\ref{fig:TEGFFs} and \ref{fig:SIGFFs} are shown both with and without the additional smoothing constraint discussed above.

\subsection{Results: Gluon GFFs}
\label{sec:results}

The procedure described in the previous section allows the determination of the $\Delta^2$-dependence of one of the eight gluon transversity GFFs, and three of the seven unpolarised gluon GFFs. At some momentum transfers, additional GFFs (or linear combinations thereof) can be determined. This study is, however, focussed on determining the $\Delta^2$-dependence of GFFs as a whole. 

The single gluon transversity GFF that can be determined, which is also the only transversity GFF that contributes in the forward limit, is shown in Fig.~\ref{fig:TEGFFs}. The forward limit of this quantity defines a transverse momentum asymmetry, which was previously determined on the same gauge ensemble as used here~\cite{Detmold:2016gpy}. While the results labelled basis 1 are more precise than those for basis 2, as there are fewer momentum and polarisation combinations that give non-zero matrix elements of basis-2 operators, they agree within uncertainties.

\begin{figure*}
	\centering
	\subfigure[Basis 1: $\tau_1^{(3)}$]{
		\includegraphics[width=0.48\textwidth]{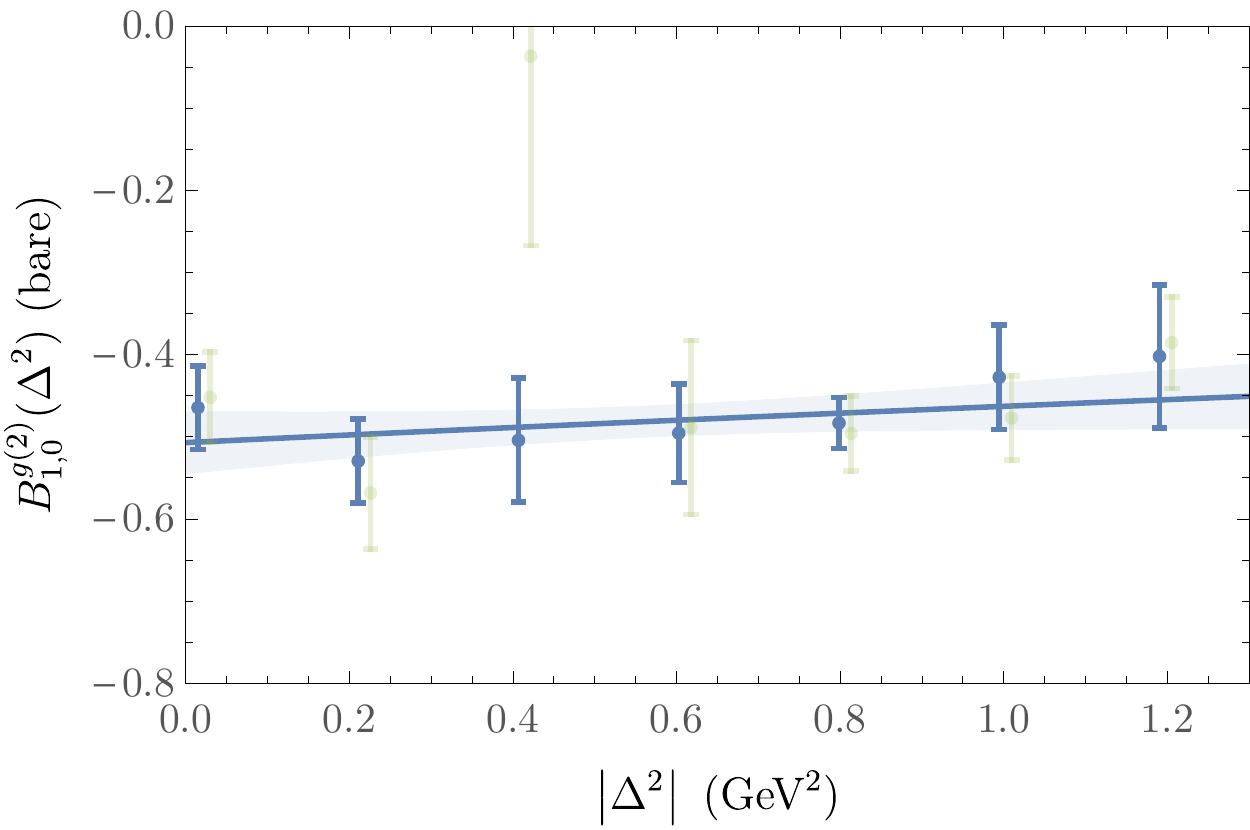}
	}
	\subfigure[Basis 2: $\tau_3^{(6)}$]{
		\includegraphics[width=0.48\textwidth]{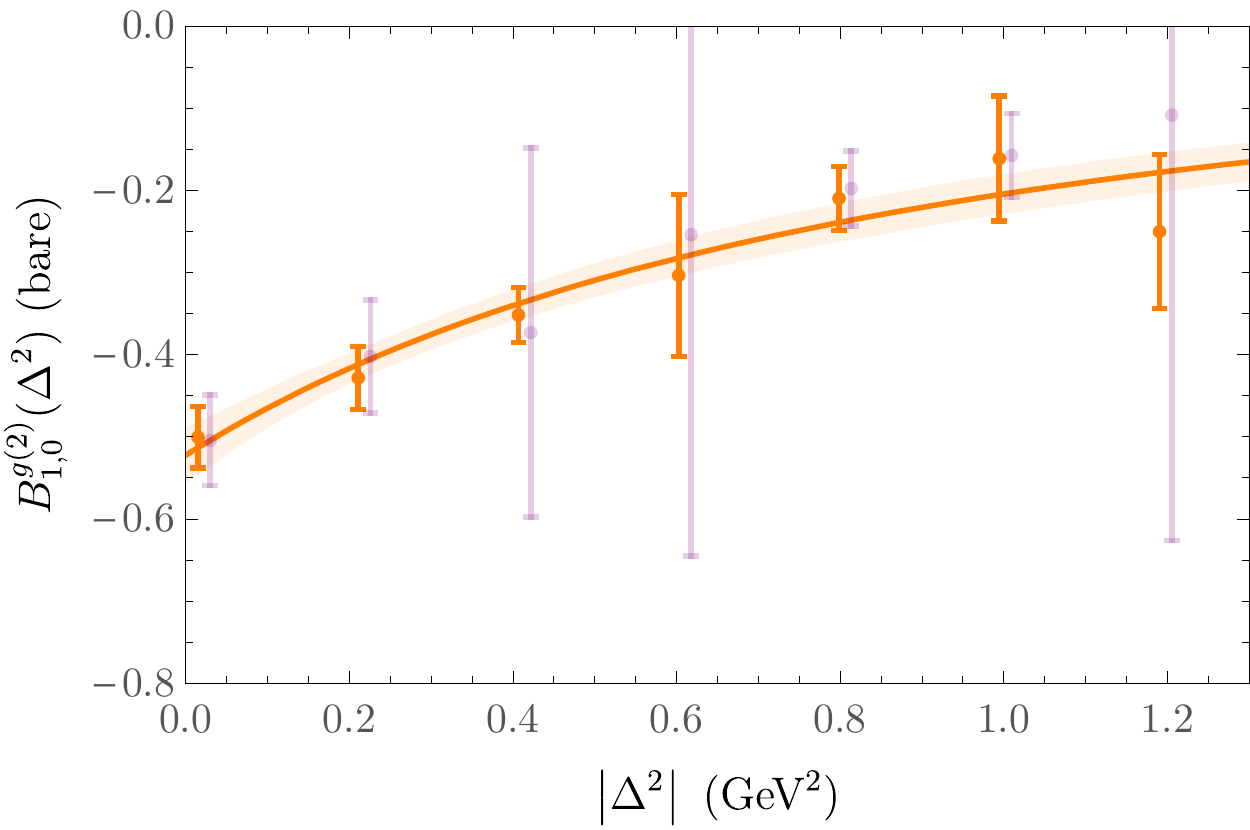}
	}
	\subfigure[Basis 1: $\tau_1^{(3)}$]{
		\includegraphics[width=0.48\textwidth]{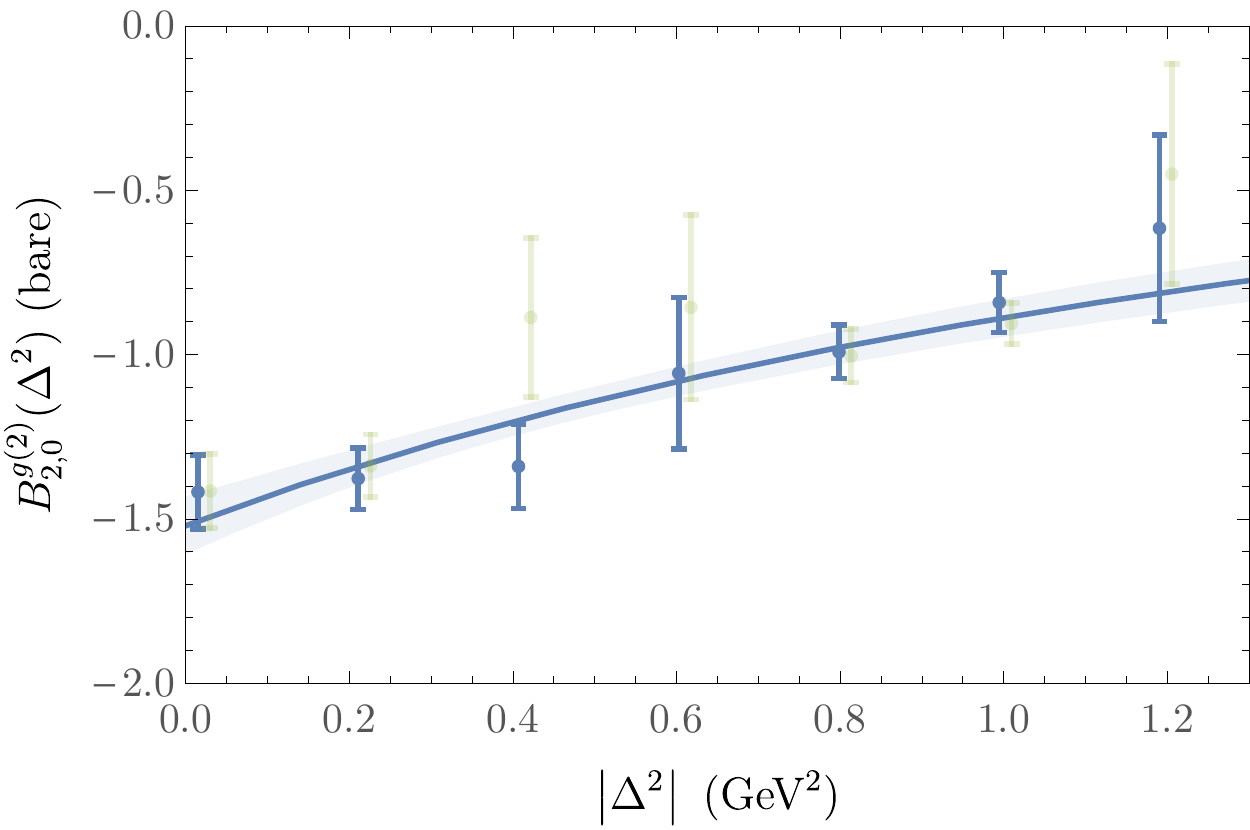}
	}
	\subfigure[Basis 2: $\tau_3^{(6)}$]{
		\includegraphics[width=0.48\textwidth]{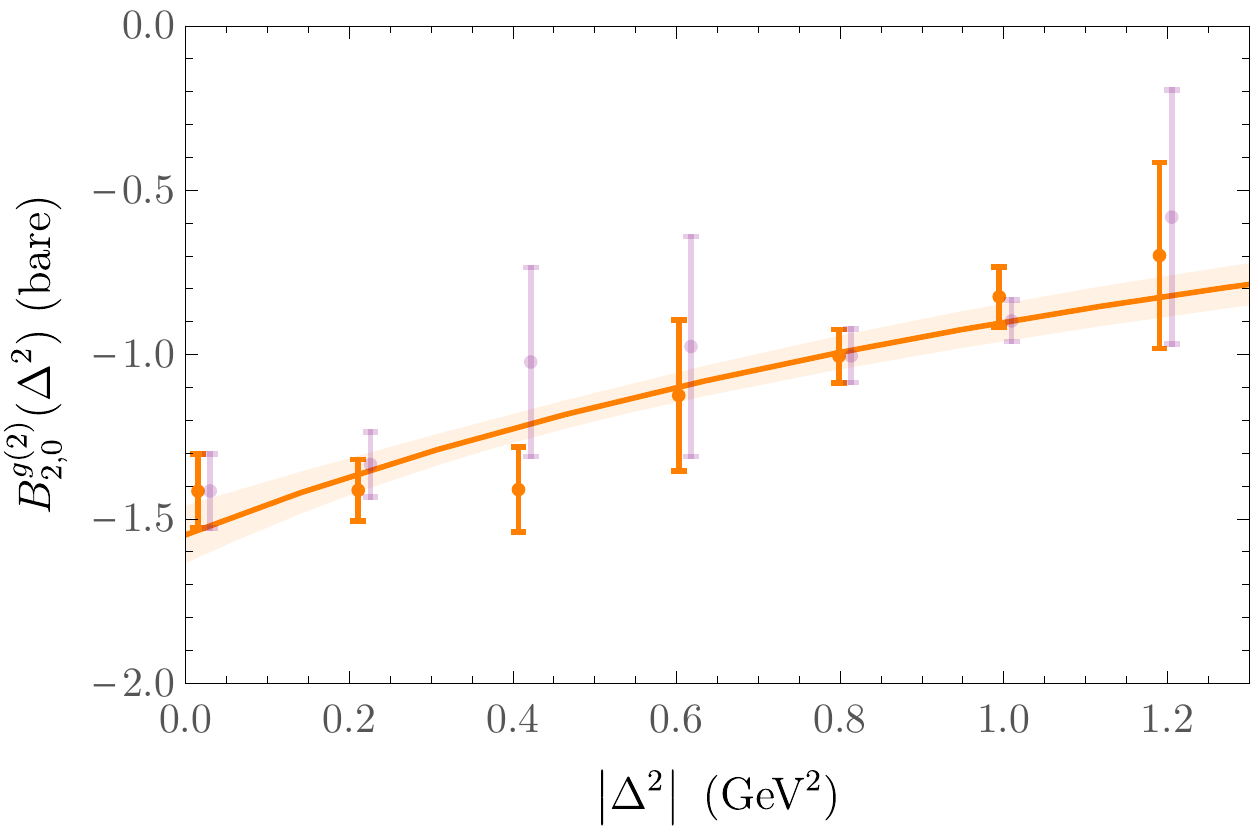}
	}
	\subfigure[Basis 1: $\tau_1^{(3)}$]{
		\includegraphics[width=0.48\textwidth]{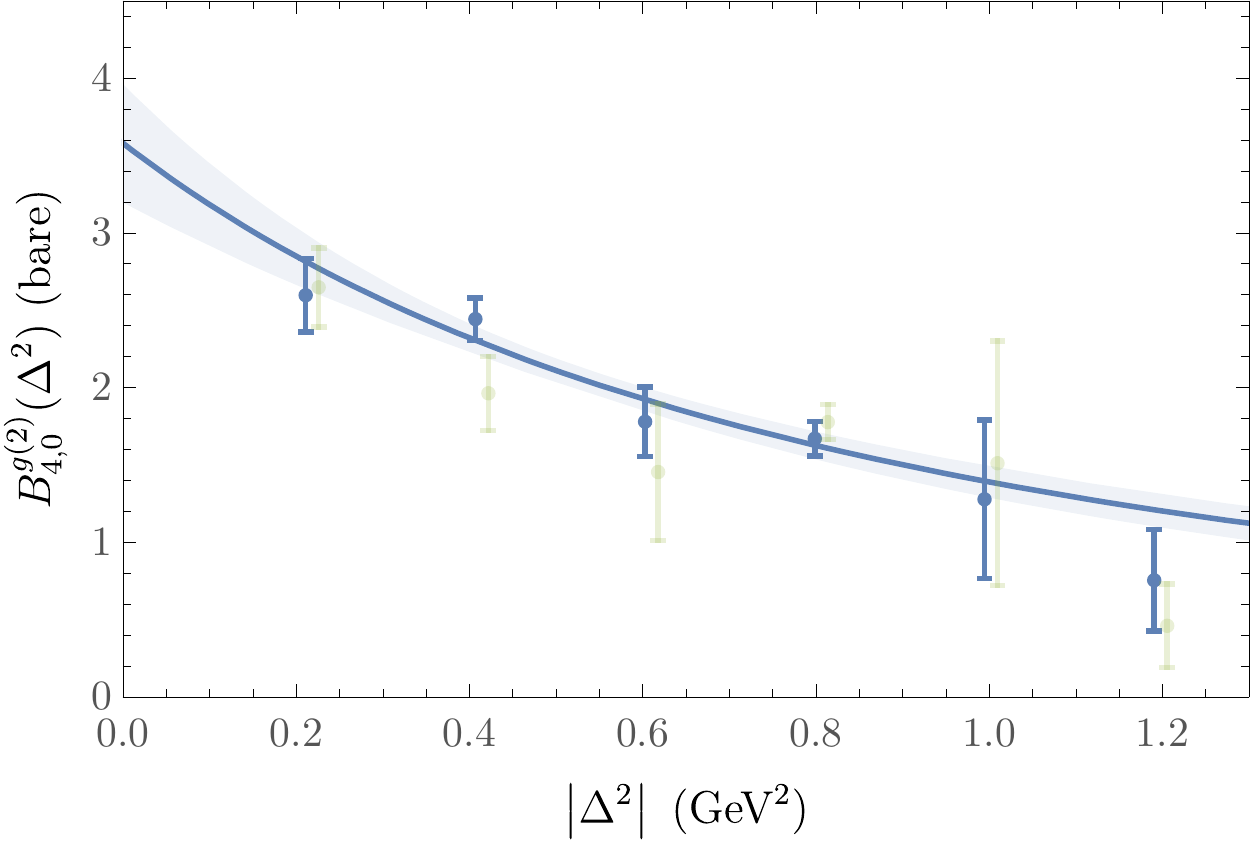}
	}
	\subfigure[Basis 2: $\tau_3^{(6)}$]{
		\includegraphics[width=0.48\textwidth]{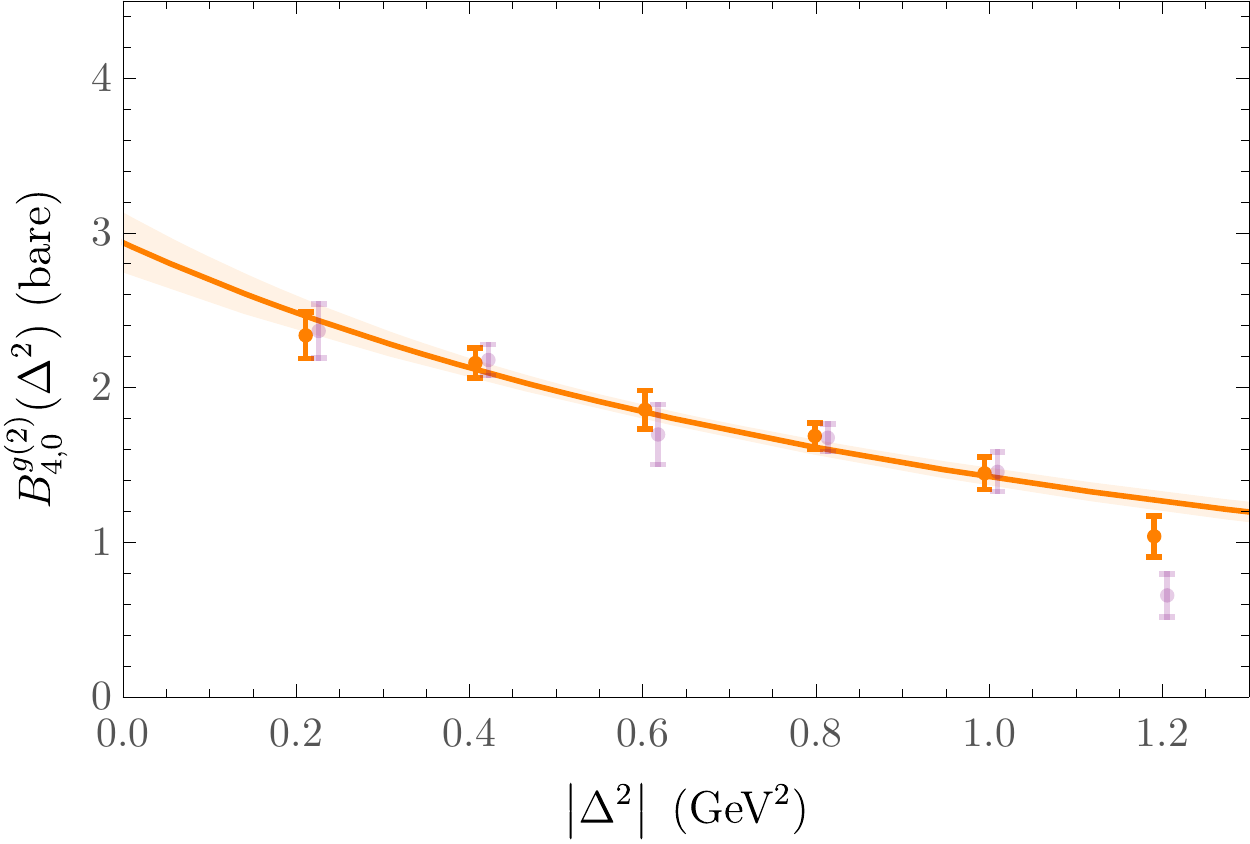}
	}
	\caption{\label{fig:SIGFFs}Spin-independent gluon GFFs determined through the analysis described in Section~\ref{sec:analysis}. Subfigures~(a), (c) and (e) show results obtained using lattice operators from the first irreducible representation considered, while the other subfigures show results from the second. As in Fig.~\ref{fig:TEGFFs}, the solid blue and orange points denote the results of the full analysis, while the faded green and purple points (offset on the horizontal axis for clarity) show the analysis repeated without the smoothness constraint discussed at the end of Section~\ref{subsec:extract}. The bands are dipole fits to the results of the full analysis against $\left|\Delta^2\right|$, shown to guide the eye.}
\end{figure*}

Results for the three spin-independent gluon GFFs that can be determined are shown in Fig.~\ref{fig:SIGFFs}. Again, the results from the two distinct bases are broadly consistent at the $1\sigma$ level, with comparable precision achieved in both data sets. While the spin-independent gluon GFFs are clearly quantitatively different from the gluon transversity GFF, this difference is as yet hard to interpret meaningfully, given the limited subset of GFFs which are determined. Within the uncertainties of this calculation, the effects of renormalisation, which have been neglected, are not significant~\cite{Alexandrou:2016ekb}.

\subsection{Quark GFFs}

To interpret the gluonic observables obtained in this study, it is interesting to compare them with the analogous quark GFFs and the electromagnetic form factors. As this first lattice determination of the gluon GFFs is performed at a single unphysical value of the quark masses, and no extrapolation to the physical point is performed, it is natural to compare gluon and quark GFFs at the same unphysical parameters. While the transversity gluon GFFs have no direct quark analogues, the spin-independent gluon GFFs have a one-to-one correspondence with the spin-independent quark GFFs. These quantities, as well as the quark electromagnetic form factors, are calculated using the same lattice setup and analysis procedures described previously, with minor differences as detailed below.

For the spin-independent strange quark GFFs and the electromagnetic form factors of the $\phi$ meson, the relevant three-point functions, $C_{jk}^\text{3pt}$ (Eq.~\eqref{eq:3pt}), are determined using the quark bilinear operators
\begin{align}\label{eq:SIopQ}
\mathcal{Q}_{\mu\nu}=&S\left[\overline{\psi} \gamma_\mu i\overleftrightarrow{D}_\nu \psi\right], \\
\mathcal{Q}_{\mu}=&\overline{\psi} \gamma_\mu \psi,
\end{align}
respectively. 
The calculations omit the disconnected couplings of the sea quarks to the currents, and $\cal {O}(\alpha)$ mixing with the gluonic operators is ignored. In addition, a single contraction is considered, so the system under consideration should be thought of as an $\overline{s}s'$ meson.

Matrix elements of the spin-independent quark operators have precisely the same form as those of the spin-independent gluon operators, given explicitly in Eq.~\eqref{eq:SIME}. The GFFs $B_{i,m}^{q(n)}$ are given the additional superscript `$q$' to identify them as the quark analogues.
The decomposition of the electromagnetic current matrix elements, which have no gluonic analogues, into form factors for spin-1 particles is~\cite{Hedditch:2007ex,Owen:2015gva}
\begin{align}
\nonumber
\left\langle p' E' \left| \overline{\psi}\right.\right. & \left.\left. \hspace{-3mm} \gamma_\mu\psi  \right| p E \right\rangle  \\\nonumber
= & \frac{\Ep{}^\alpha E^\beta}{2 \sqrt{\mathcal{E}\mathcal{E}'}}\left( \vphantom{\frac{1}{M^2}}-2G_1(\Delta^2)g_{\alpha\beta}P_\mu \right.\\\nonumber
&\hspace{1.5cm}- \left.G_2(\Delta^2)\left(\Delta_\beta g_{\mu\alpha} - \Delta_\alpha g_{\mu\beta} \right) \right.\\\label{eq:EMFFs}
&\hspace{1.5cm}+ \left.\frac{1}{M^2}G_3(\Delta^2)\Delta_\alpha\Delta_\beta P_\mu\right),
\end{align}
where, as before, $P=(p+p')/2$ is the average momentum, and the momentum transfer is defined as $\Delta = p'-p$. The initial and final-state energies are $\mathcal{E}$ and $\mathcal{E}'$ (and the polarisations, as before, are $E$ and $E'$).

The three-point functions needed for the LQCD determination of these matrix elements are constructed using sequential propagators with fixed sink momentum $\vec{p}\,'=0$, since each additional momentum used carries additional computational cost. As a result, there is less information (in the form of fewer independent equations in the system that determines the GFFs or FFs) than in the gluonic study discussed above.
Moreover, unlike in the gluon case, the construction of three-point functions with different sink times requires separate sequential propagator computation. Three sink times, $t_\text{sink}\in\{12,14,16\}$ in lattice units, are used here.
The statistical behaviour of these matrix elements is, however, less noisy than that of the gluon operator correlators.

As was discussed previously for the gluon case, calculations are performed using lattice operators which transform irreducibly under the hypercubic group H(4). 
The vector current is implemented using the naive local current, related to the continuum current by a renormalisation factor $Z_V$ that is determined by demanding unit charge. This transforms in the $\tau_1^{(4)}$ representation.
For the spin-independent quark operator in Eq.~\eqref{eq:SIopQ}, the irreducible representations have exactly the same form as those for the spin-independent gluon operator, which are given explicitly in Appendix~\ref{app:latticeOps}. A single lattice representation of the spin-independent quark operator is used, namely $\overline{Q}^{E}_{2,14}$, analogous to the gluon quantity labelled as $\overline{O}^{E}_{2,14}$ in Appendix~\ref{app:latticeOps}.

Given the three point functions constructed as described above, the analysis to extract the spin-independent quark GFFs proceeds in the same way as was detailed for the spin-independent gluon GFFs above. The only modification is that, with only three sink times available, identified plateaus in the $(t$--$\tau)$ plane are considered acceptable if they include a minimum of 5 timeslices.
For the electromagnetic form factors, where a complete extraction of all FFs is possible, both this method (with all three FFs considered `dominant') and a separate analysis, in which ratios of two and three-point functions are combined to give direct extractions of the individual FFs, are used. This second method is discussed in detail in Appendix~\ref{app:bob} and gives results consistent with the more general method that must be used in the cases where larger numbers of form factors contribute. This comparison also gives confidence that this latter method is reliable.

\begin{figure}
	\centering
	\subfigure[]{
		\includegraphics[width=0.48\textwidth]{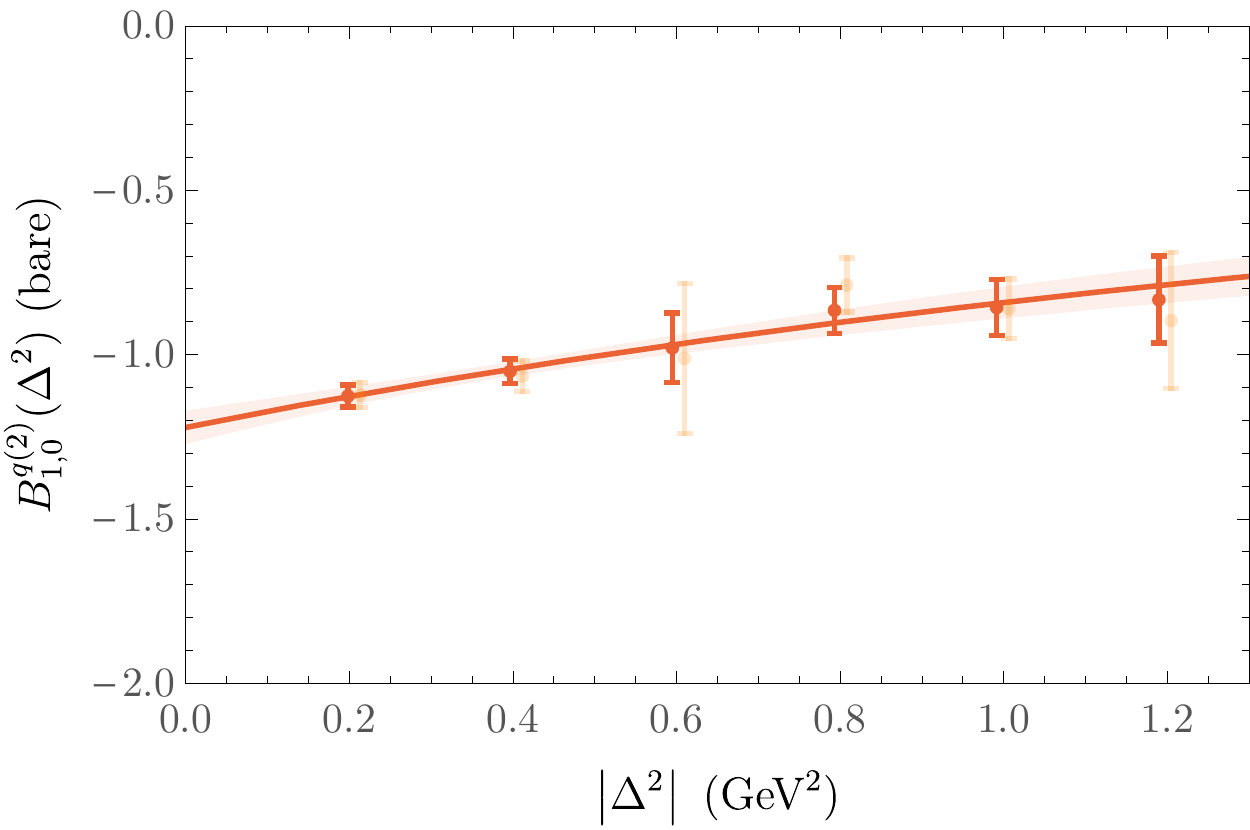}
	}
	\subfigure[]{
	\includegraphics[width=0.48\textwidth]{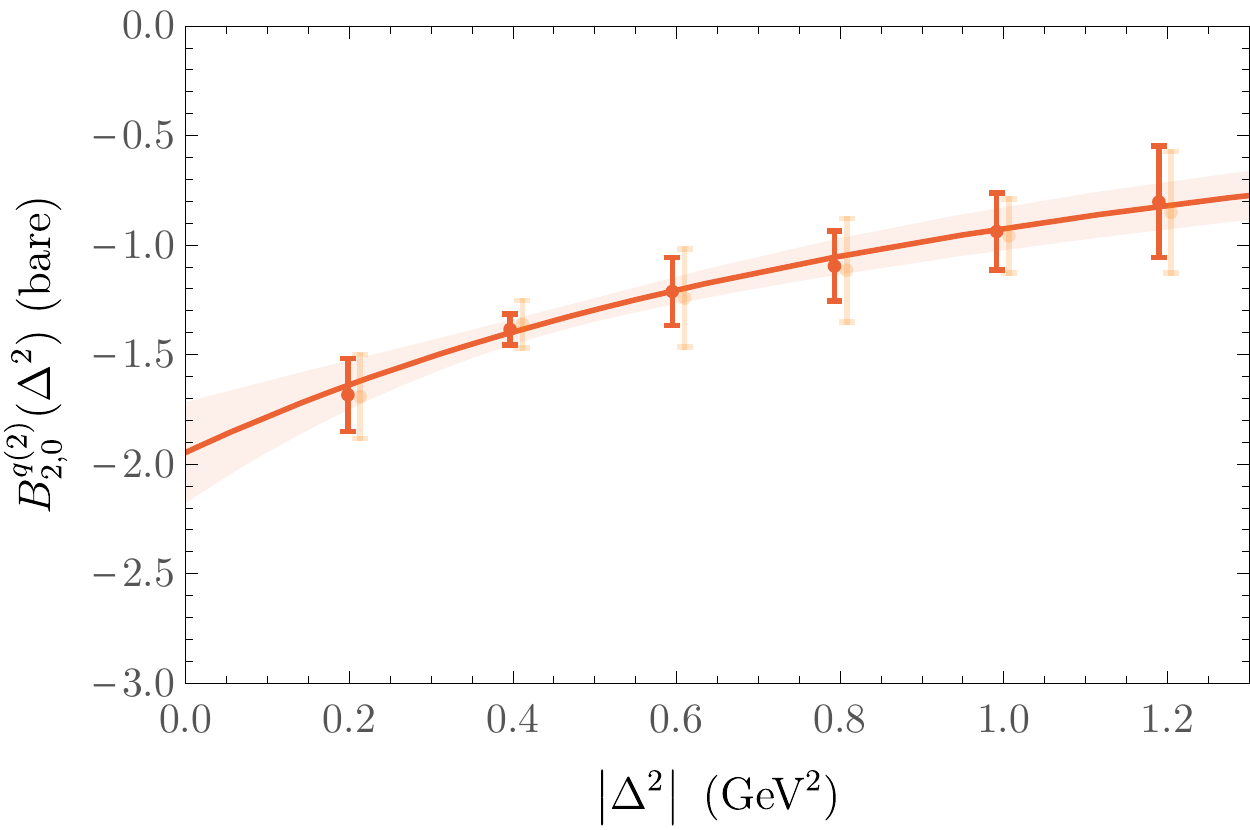}
}
	\subfigure[]{
		\includegraphics[width=0.48\textwidth]{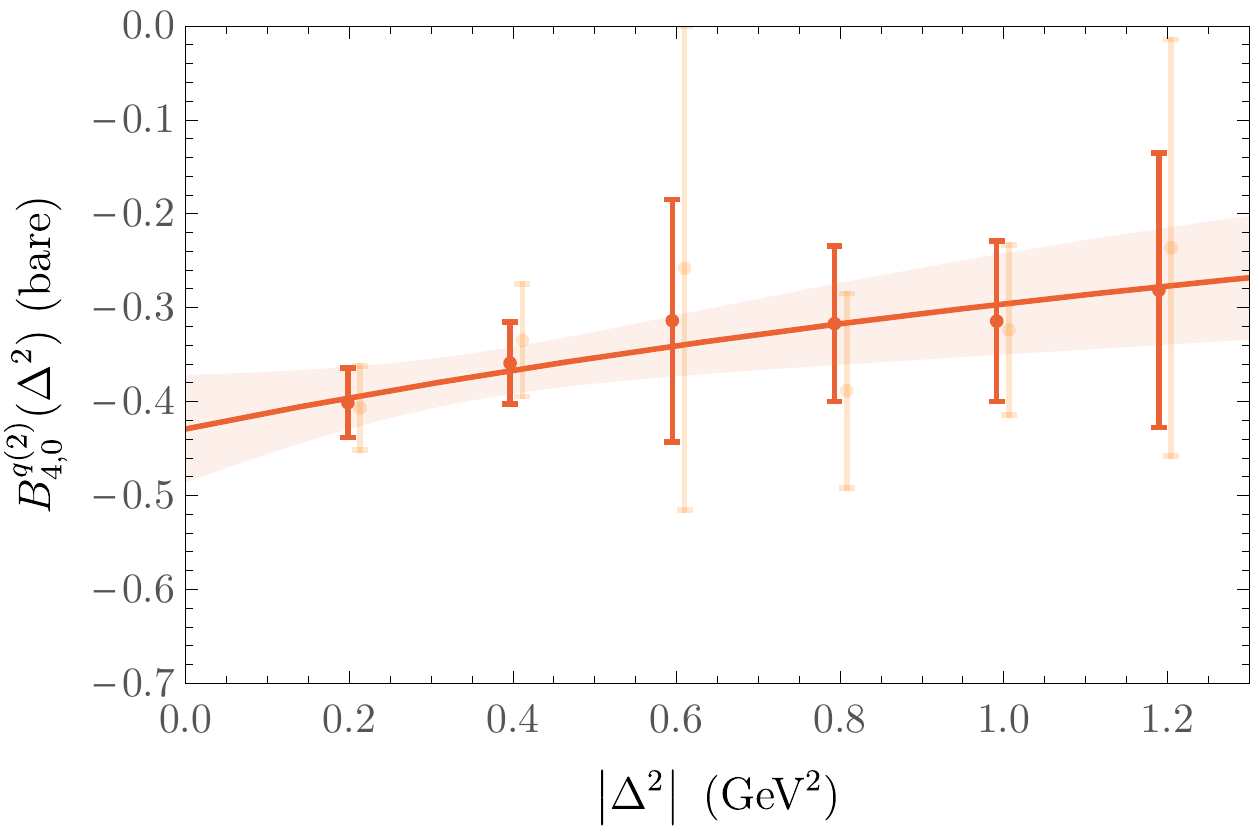}
	}
	\caption{\label{fig:QGFFs}Spin-independent quark GFFs determined as described in the text. As in Figs.~\ref{fig:TEGFFs} and \ref{fig:SIGFFs}, the solid red points denote the results of the full analysis, while the faded orange points (offset on the horizontal axis for clarity) show results obtained without the smoothness condition discussed at the end of Section~\ref{subsec:extract}. The bands are dipole fits to the results of the full analysis against $\left|\Delta^2\right|$.}
\end{figure}

\begin{figure}
	\centering
	\subfigure[]{
		\includegraphics[width=0.48\textwidth]{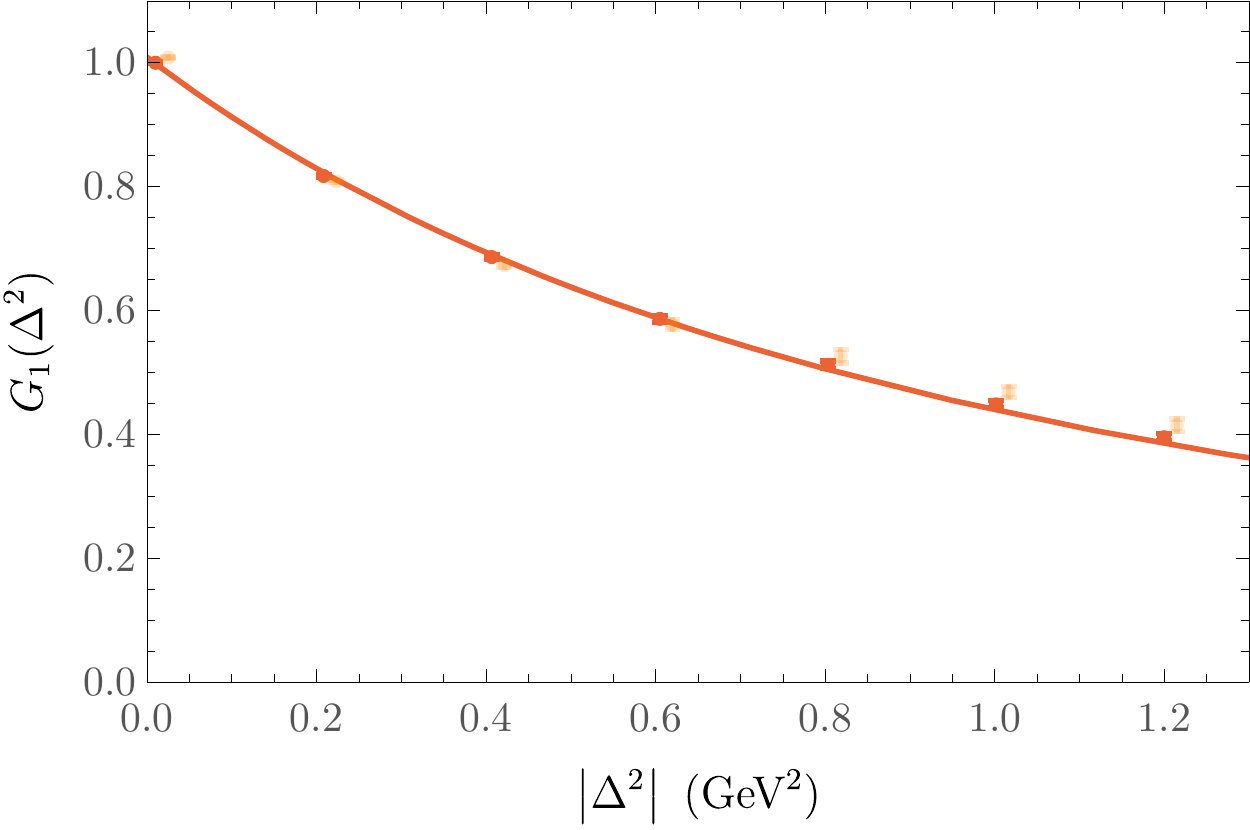}
	}
	\subfigure[]{
		\includegraphics[width=0.48\textwidth]{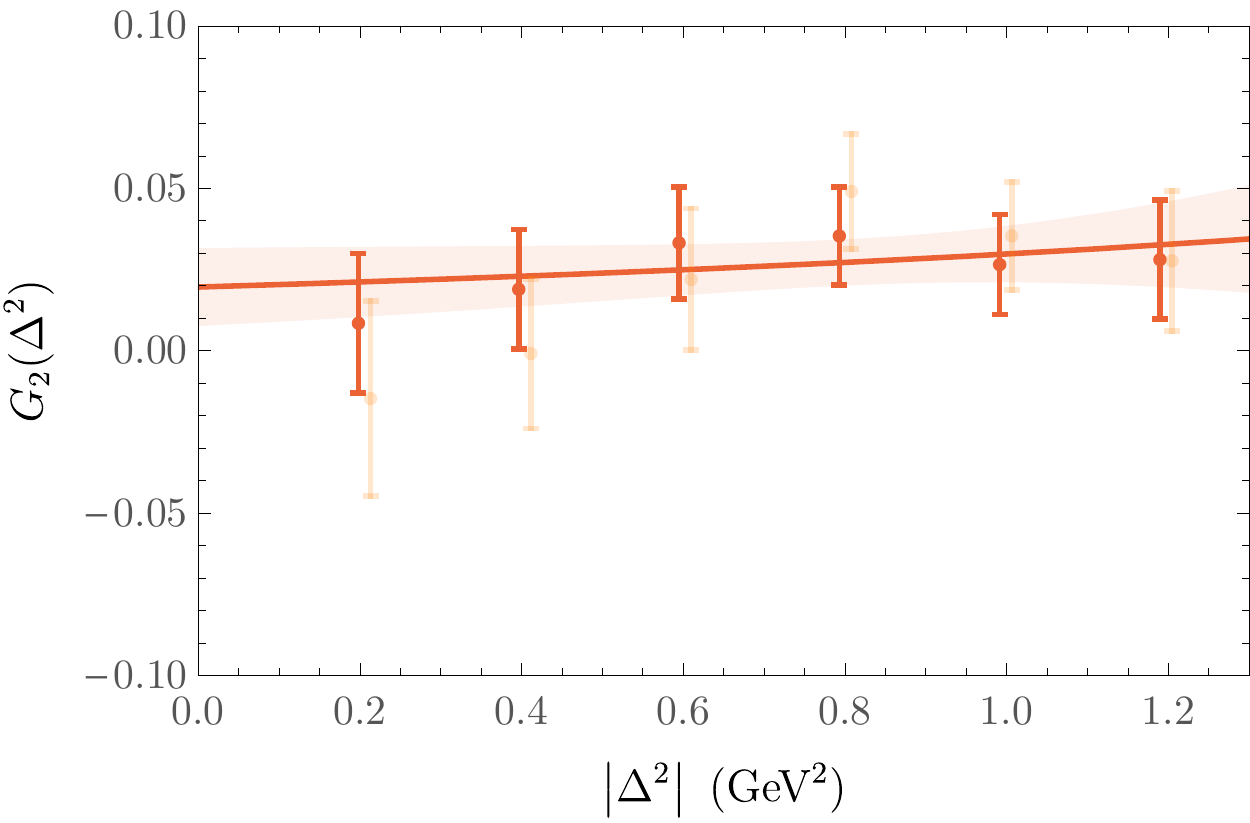}
	}
	\subfigure[]{
		\includegraphics[width=0.48\textwidth]{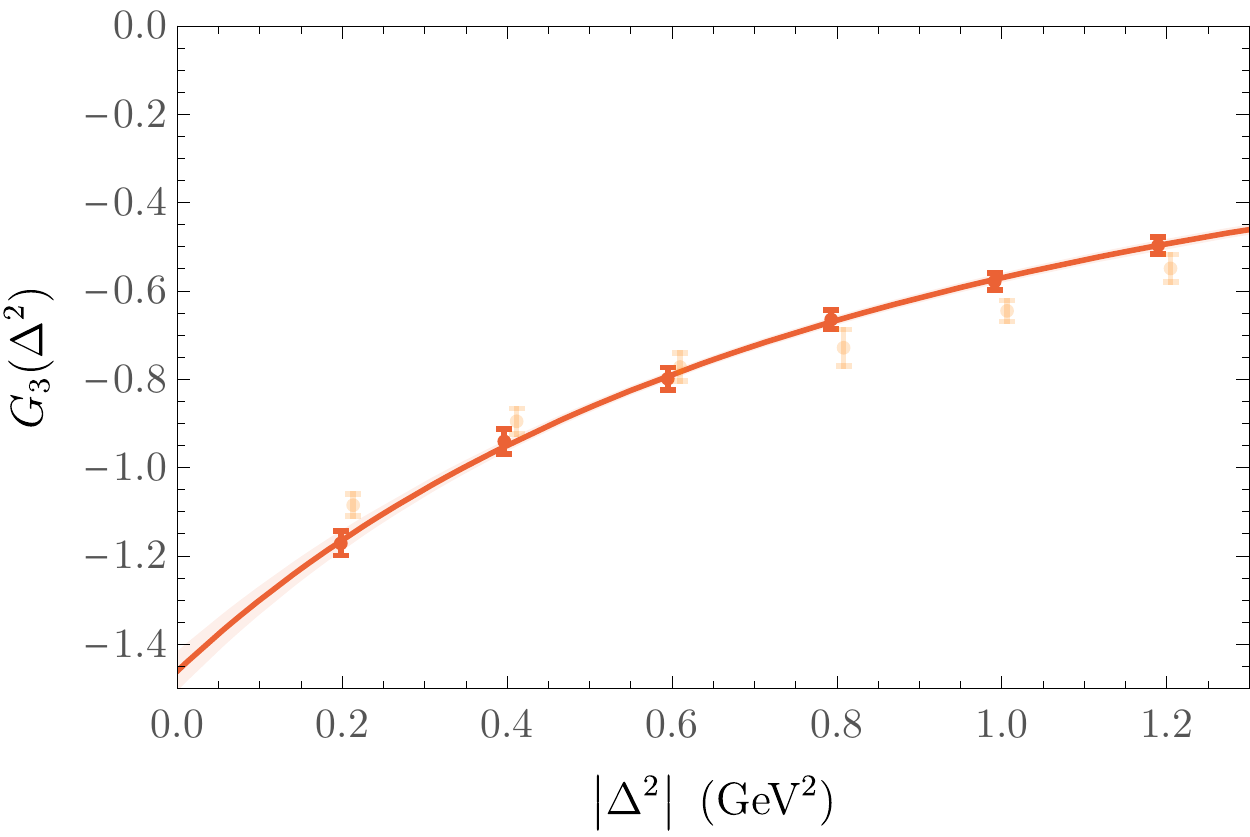}
	}
	\caption{\label{fig:QFFs}Quark electromagnetic form factors determined as described in the text. As in Figs.~\ref{fig:TEGFFs} and \ref{fig:SIGFFs}, the solid red points denote the results of the full analysis, while the faded orange points (offset on the horizontal axis for clarity) show results obtained without application of the smoothness condition discussed at the end of Section~\ref{subsec:extract}. The bands are dipole fits to the results of the full analysis against $\left|\Delta^2\right|$.}
\end{figure}

This procedure allows the determination of the $\Delta^2$-dependence of three of the seven unpolarised quark GFFs. These three quantities are the direct analogues of the three spin-independent gluon GFFs that were determined as described in the previous sections, and can therefore be compared with these one-to-one. The results are shown in Fig.~\ref{fig:QGFFs}. The forward-limit quantities $B_{1,0}^{q(2)}(0)$ and $B_{2,0}^{q(2)}(0)$ satisfy the Soffer-type bounds derived on their relation in Ref.~\cite{Bacchetta:2001rb}. 
These quantities were previous calculated for a heavy $\rho$ meson in a quenched calculation~\cite{Best:1997qp}. Translating the results presented here to the notation in that work, the linear combination of $B_{1,0}^{q(2)}(0)$ and $B_{2,0}^{q(2)}(0)$ named $a_1$ has a comparable value, but the combination $d_1$ has a different sign and magnitude, possibly due to the effects of unquenching.
The results for the form factors of the vector current are shown in Fig.~\ref{fig:QFFs} (and with a different choice of decomposition in Fig.~\ref{fig:QFFsCQM}). As discussed above, in this case all three form factors can be determined.

\section{Discussion}

\begin{figure}
	\centering
	\includegraphics[width=0.48\textwidth]{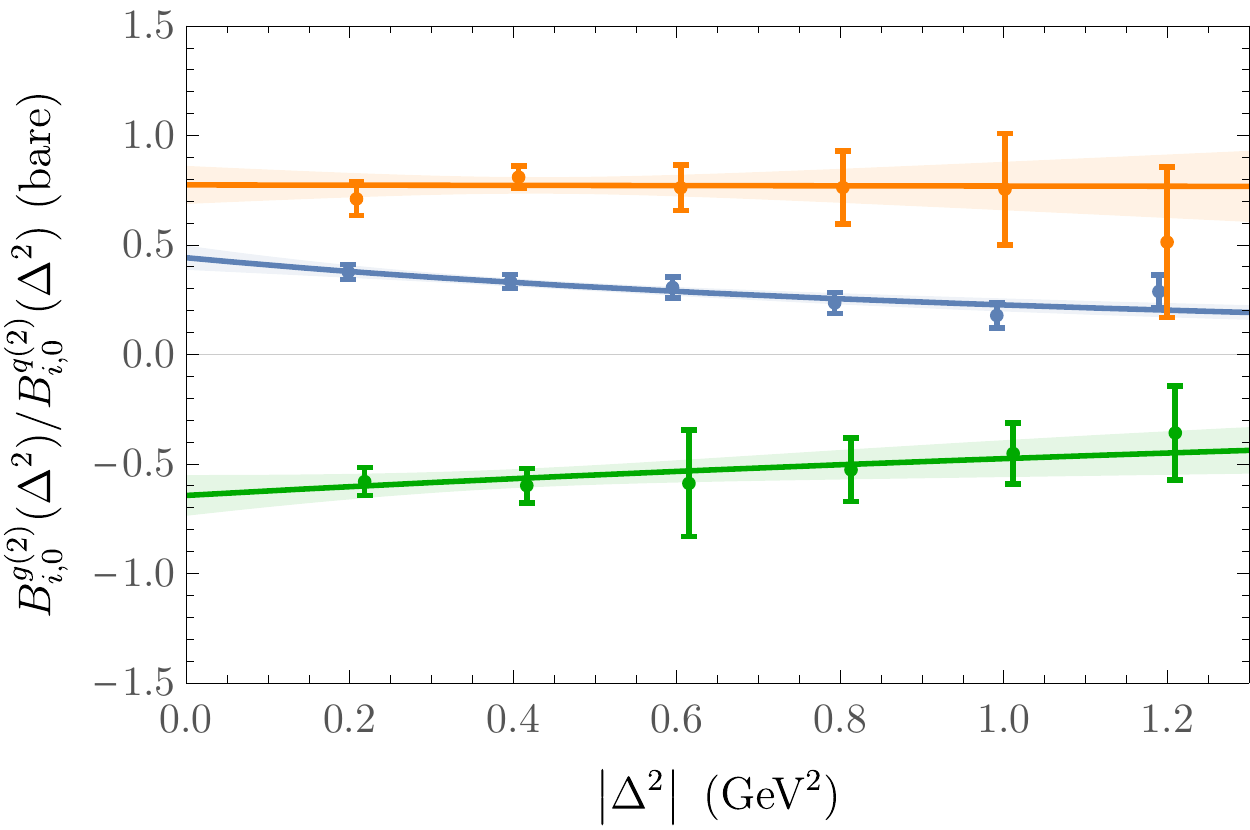}
	\caption{\label{fig:gqrat} Ratio of gluon to quark spin-independent GFFs, determined from the full analysis described in the text. The blue (middle), orange (top) and green (bottom) sets of results correspond to $i=\{1,2,4\}$, respectively. The bottom data set, for the ratio $B_{4,0}^{g(2)}(0)/B_{4,0}^{q(2)}$, is rescaled by a factor of $1/10$. The bands are dipole fits to the results of the full analysis against $\left|\Delta^2\right|$.}
\end{figure}

While many observables related to the quark structure of hadrons and nuclei have now been both measured experimentally and calculated at some level from QCD, including electromagnetic form factors and quark momentum distributions, the gluon structure of these particles remains far more mysterious.
One question of fundamental interest is the spatial distributions of gluons relative to that of quarks; is the `gluonic radius' of a hadron larger, smaller, or of a similar size to the corresponding quark radius. This, while interesting, is a somewhat nebulous question. 
The observables considered in this work, namely the spin-independent and transversity gluon GFFs, each define a gluonic radius, each of which could, in principle, be different, and there is no unique basis for the GFF decomposition. The spin-independent gluon GFFs have direct quark analogues and can be compared with these on a one-to-one basis. 
While there are clear quantitative differences between each of the the quark and gluon GFFs in the three matched pairs that were studied here, these differences vary (i.e., the spin-independent quark and gluon GFFs are not universally related by an approximate scaling or sign change, as shown in Fig.~\ref{fig:gqrat}). The gluon transversity distribution is, again, quantitatively different, but this difference is hard to interpret in a physical sense. Resolving a full three-dimensional picture of the gluon structure of the $\phi$ meson will require more precise calculations that are able to resolve the entire basis of GFFs for the first moments of the gluon distributions, and also extend to higher moments.

An EIC will, for the first time, allow experimental measurements of gluon GFFs in nucleons and nuclei. The present work represents a demonstration that QCD predictions of gluonic structure quantities can be obtained using LQCD. Future LQCD studies of nucleons and nuclei with fully-controlled uncertainties will inform the design and targets of an EIC experimental program, guide the interpretation of first measurements, and, for some quantities, act as theory benchmarks for the EIC.

\section*{Acknowledgements}

This work is supported by the U.S. Department of Energy under Early Career Research Award DE-SC0010495 and under Grant No. DE-SC0011090. The work of WD is supported in part by the U.S. Department of Energy, Office of Science, Office of Nuclear Physics, within the framework of the TMD Topical Collaboration. The Chroma software library~\cite{Edwards:2004sx} was used in the data analysis. We thank Ian Clo\"et, Bob Jaffe, James Maxwell, Richard Milner, and Ross Young for helpful discussions, and Kostas Orginos for the production of the gauge configurations used in this work. We thank W. Cosyn, A. Freese and B. Pire for bringing errors in a previous version of this work to our attention.

\appendix

\section{Explicit expressions for lattice operators}
\label{app:latticeOps}

Bases of Euclidean operators that transform irreducibly under the hypercubic group H(4) are defined, for different symmetry classes of operators, in Ref.~\cite{Gockeler:1996mu}. The same notation as in that work is used here, with irreducible representations labelled by $\tau_i^{(n)}$ where $n$ denotes the dimension of the representation and $i$ enumerates representations.

The Euclidean analogue of the transversity operator defined in Eq.~\eqref{eq:TTop}, for $n=2$, is built from the quantity
\begin{equation}
\mathcal{O}^{(E)}_{\mu\nu\mu_1 \mu_{2}} = G^{(E)}_{\mu\mu_1}G^{(E)}_{\nu \mu_2},
\end{equation}
where the clover definition of the gluon field strength tensor is used for $G_{\mu\nu}^{(E)}$ in the numerical calculations.
While there are three irreducible representations that are safe from mixing with lower or same-dimensional quark bilinear operators, namely $\tau_1^{(2)}$, $\tau_2^{(6)}$ and $\tau_2^{(2)}$~\cite{Detmold:2016gpy}, only a subset of the basis operators of these representations, which provide the cleanest signals, are considered here. In particular, three of the six basis operators from $\tau_2^{(6)}$, and both basis operators from $\tau_2^{(2)}$, have vanishing matrix elements in states that have small momenta in lattice units (for example, with zero momentum transfer, these operators only have nonvanishing matrix elements for boosts with $\vec{p}\,^2\ge 3$~\cite{Detmold:2016gpy}). The five remaining operators from representations $\tau_1^{(2)}$ and $\tau_2^{(6)}$ are non-vanishing in a larger number of external states, and so provide the most information. These are considered in this work. 

For $\tau_1^{(2)}$, the basis vectors are~\cite{Gockeler:1996mu,Detmold:2016gpy}:
\begin{align}\nonumber
\mathcal{O}_{{1,1}}^{(E)} = & \frac{1}{8\sqrt{3}}\left(-2 {\mathcal{O}}^{(E)}_{1122}+{\mathcal{O}}^{(E)}_{1133}+{\mathcal{O}}^{(E)}_{1144} \right.\\
& \left. \hphantom{\frac{1}{8\sqrt{3}}\,\,\,\,\,} +{\mathcal{O}}^{(E)}_{2233}+{\mathcal{O}}^{(E)}_{2244}-2 {\mathcal{O}}^{(E)}_{3344} \right), \\
\mathcal{O}_{{1,2}}^{(E)} = & \frac{1}{8} \left({\mathcal{O}}^{(E)}_{1144}+{\mathcal{O}}^{(E)}_{2233}-{\mathcal{O}}^{(E)}_{1133}-{\mathcal{O}}^{(E)}_{2244}\right).
\end{align}

The $\tau_2^{(6)}$ vectors which are used here are:
\begin{align}
\mathcal{O}_{{2,2}}^{(E)} = & \frac{1}{4} \left({\mathcal{O}}^{(E)}_{1124}+{\mathcal{O}}^{(E)}_{2334}\right),\\
\mathcal{O}_{{2,4}}^{(E)} = & \frac{1}{4} \left({\mathcal{O}}^{(E)}_{1224}-{\mathcal{O}}^{(E)}_{1334}\right),\\
\mathcal{O}_{{2,5}}^{(E)} = & \frac{1}{4} \left({\mathcal{O}}^{(E)}_{1134}-{\mathcal{O}}^{(E)}_{2234}\right).
\end{align}

The Euclidean analogue of the spin-independent gluonic operator in Eq.~\eqref{eq:SIop}, for $n=0$, is constructed from
\begin{equation}
\overline{\mathcal{O}}^{(E)}_{\mu_1\mu_2}=G_{\mu_1\alpha}^{(E)}G_{\mu_2\alpha}^{(E)}.
\end{equation}
Basis operators from two irreducible representations are considered here. 
A basis of operators for the $\tau_1^{(3)}$ representation is:
\begin{align}\label{eq:myfirst}
\overline{\mathcal{O}}_{{1,1}}^{(E)} = & \frac{1}{2}\left(\overline{\mathcal{O}}^{(E)}_{11}+\overline{\mathcal{O}}^{(E)}_{22}-\overline{\mathcal{O}}^{(E)}_{33}-\overline{\mathcal{O}}^{(E)}_{44}\right), \\
\overline{\mathcal{O}}_{{1,2}}^{(E)} = & \frac{1}{\sqrt{2}} \left(\overline{\mathcal{O}}^{(E)}_{33}-\overline{\mathcal{O}}^{(E)}_{44}\right),\\
\overline{\mathcal{O}}_{{1,3}}^{(E)} = & \frac{1}{\sqrt{2}} \left(\overline{\mathcal{O}}^{(E)}_{11}-\overline{\mathcal{O}}^{(E)}_{22}\right).
\end{align}
For $\tau_3^{(6)}$ the vectors are:
\begin{align}\label{eq:mylast}
\overline{\mathcal{O}}_{{2,\mu\nu}}^{(E)} = & \frac{1}{\sqrt{2}} \left(\overline{\mathcal{O}}^{(E)}_{\mu\nu}+\overline{\mathcal{O}}^{(E)}_{\nu\mu}\right), \hspace{2mm} 1\le \mu < \nu \le 4,
\end{align}
where only the operators with $\nu=4$ are used here, as they provide the cleanest signals at most momenta.

The Minkowski-space analogue of each basis operator is determined by applying the relations
\begin{align}
G^{(E)}_{ij} & = G_{ij} \text{   if } i,j\in \{1,2,3\},\\
G^{(E)}_{4j} & = (-i)G_{0j},
\end{align}
to the Euclidean-space form. These Minkowski operators are used, as described in Sec.~\ref{sec:MEs}, to match the numerical LQCD results for the matrix elements of operators to the expressions for these quantities in terms of GFFs. 

Explicit expressions for the Euclidean and Minkowski-space quark operators are given in Ref.~\cite{Gockeler:1995wg}. The structure of H(4) irreducible representations constructed from the Euclidean operators is identical to that given in Eqs.~\eqref{eq:myfirst} through \eqref{eq:mylast} for the gluon case.

\pagebreak
\begin{widetext}

\section{GFFs of the spin-dependent gluon operator in spin-1 states}
\label{app:GFFs}

Although the spin-dependent gluon distributions are not studied numerically in this work, the GFF decomposition of matrix elements of the spin-dependent gluon operator in Eq.~\eqref{eq:SDop} is derived here. It can be expressed as (equivalent to the decomposition in Ref.~\cite{Cosyn:2018thq} after application of the Schouten identities)
\begin{align*}
\left\langle p' E'\left|S\left[\tilde{G}_{\mu\alpha}\hspace{-3mm}\right.\right.\right.&\left.\left.\left. i \overleftrightarrow{D}_{\mu_1}\ldots i\overleftrightarrow{D}_{\mu_n}G_{\mu_{n+1}}^{\,\,\alpha}\right]\right|p E\right\rangle \\
=  \sum_{\substack{m \text{ even}\\m=0}}^{n+1} & \tilde{B}^{(n+2)}_{1,m}(\Delta^2)\,S\left[\epsilon_{\mu\alpha\beta\gamma}E^\alpha\Ep{}^\beta P^\gamma  \DD_{\mu_1}\ldots\DD_{\mu_{m}}P_{\mu_{m+1}}\ldots P_{\mu_{n+1}} \right]\\
& + \sum_{\substack{m \text{ even}\\m=0}}^{n}
\tilde{B}^{(n+2)}_{2,m}(\Delta^2)\,S\left[\epsilon_{\mu\alpha\beta\gamma}E^\alpha\Ep{}^\beta \DD^\gamma  \DD_{\mu_1}\ldots\DD_{\mu_{m+1}}P_{\mu_{m+2}}\ldots P_{\mu_{n+1}} \right]\\
& + \sum_{\substack{m \text{ even}\\m=0}}^{n}
\tilde{B}^{(n+2)}_{3,m}(\Delta^2)\,S\left[\left(\epsilon_{\mu\alpha\beta\gamma}E^\alpha P^\beta \DD^\gamma \Ep_{\mu_{n+1}} + \epsilon_{\mu\alpha\beta\gamma}\Ep{}^\alpha P^\beta \DD^\gamma E_{\mu_{n+1}} \right)\DD_{\mu_1}\ldots\DD_{\mu_{m}}P_{\mu_{m+1}}\ldots P_{\mu_n} \right]\\
& + \sum_{\substack{m \text{ even}\\m=0}}^{n+1} \frac{\tilde{B}^{(n+2)}_{4,m}(\Delta^2)}{M^2}S\left[\epsilon_{\alpha\beta\gamma\delta}E^\alpha\Ep{}^\beta P^\gamma \DD^\delta \DD_\mu \DD_{\mu_1}\ldots\DD_{\mu_{m}}P_{\mu_{m+1}}\ldots P_{\mu_{n+1}} \right].
\end{align*}
Here $S$ denotes symmetrisation and trace-subtraction in all free indices. The average momentum is defined as $P=(p+p')/2$, and the momentum transfer is $\DD=p'-p$. Note that, because of the symmetries of the operator, $\tilde{B}_{i,0}^{(2)}(\Delta^2)=0$ for all $i\in \{1\ldots 7\}$. Furthermore, none of the GFFs $\tilde{B}_{i,m}^{(n)}(\Delta^2)$ contribute to forward-limit matrix elements, which must vanish.

\end{widetext}

\section{Example of a system of linear equations for the GFFs}
\label{app:lineqs}

This section gives an explicit example of the systems of equations that are solved to extract the GFFs as described in Section~\ref{subsec:extract}. The case shown is for the gluon transversity operator in the first basis ($\tau_1^{(2)}$), at the first non-zero momentum transfer. This is the simplest case at non-zero momentum, as this basis includes only two operators.

Matching plateau-fits to the ratios $R_{jk}(\vec{p},\vec{p}\,',t,\tau,\mathcal{O})$ (Eq.~\eqref{eq:ratR}) to linear combinations of the GFFs $A_{i,0}^{(2)}(\Delta^2=1)$ as described in Section~\ref{subsec:extract}, for all choices of polarisation and momenta (up to $\vec{p}\,^2=4$ and $\vec{p}\,'^2=4$) that give $\Delta^2=1$ in lattice units and for all transversity operators in basis 1 ($\tau_1^{(2)}$), defines 154 linear equations. This system is reduced to 39 equations by averaging the ratios (before fitting in $t$ and $\tau$) over all choices of momentum and polarisation that give the same linear combination of GFFs up to a sign. The polarisations and momenta defining one member of each reduced set are given in Table~\ref{tab:momenta}.

\begin{table}[!b]
	{\small
		\begin{minipage}{0.48\columnwidth}
	\begin{tabular}{cccc}\toprule
		$\vec{p}$& $\vec{p}\,'$ & $E$ & $E'$ \\ \toprule
		~[1, 0, 0] &[1, -1, 0] & 3 & 3 \\
		~[0, 1, 0] &[0, 1, -1] & 3 & 3 \\ 
		~[0, 0, 0] &[0, -1, 0] & 3 & 3 \\ 
		~[0, 0, 1] &[0, -1, 1] & 3 & 3 \\ 
		~[0, 0, 0] &[0, 0, -1] & 3 & 3 \\ 
		~[1, 0, 0] &[1, -1, 0] & 2 & 2 \\ 
		~[0, 1, 0] &[0, 1, -1] & 1 & 1 \\ 
		~[0, 0, 1] &[0, -1, 1] & 1 & 1 \\ 
		~[0, 0, 0] &[0, -1, 0] & 1 & 1 \\ 
		~[1, 0, 0] &[1, -1, 0] & 1 & 1 \\ 
		~[0, 0, 1] &[0, -1, 1] & 2 & 2 \\ 
		~[0, 0, 0] &[0, -1, 0] & 2 & 2 \\ 
		~[0, 1, 0] &[0, 1, -1] & 2 & 2 \\ 
		~[0, 0, 0] &[0, 0, -1] & 2 & 2 \\ 
		~[1, 0, 0] &[1, -1, 0] & 2 & 1 \\ 
		~[0, 0, 1] &[0, -1, 1] & 2 & 3 \\
		~[0, 1, 0] &[0, 1, -1] & 2 & 3 \\ 
		~[0, 0, 1] &[0, -1, 1] & 3 & 2 \\ 
		~[1, 0, 0] &[1, -1, 0] & 1 & 2 \\
		~[0, 1, 0] &[0, 1, -1] & 3 & 2 \\\hline
	\end{tabular}
\end{minipage}
\begin{minipage}{0.48\columnwidth}
\begin{tabular}{cccc}\toprule
	$\vec{p}$& $\vec{p}\,'$ & $E$ & $E'$ \\ \toprule
~[0, 0, 1] &[0, -1, 1] & 1 & 1 \\
~[0, 0, 1] &[0, -1, 1] & 2 & 2 \\
~[0, 1, 0] &[0, 1, -1] & 1 & 1 \\
~[1, 0, 0] &[1, -1, 0] & 2 & 2 \\
~[0, 0, 0] &[0, 0, -1] & 2 & 2 \\
~[0, 1, 0] &[0, 1, -1] & 2 & 2 \\
~[0, 0, 0] &[0, -1, 0] & 1 & 1 \\
~[0, 0, 0] &[0, -1, 0] & 2 & 2 \\
~[1, 0, 0] &[1, -1, 0] & 1 & 1 \\
~[0, 1, 0] &[0, 1, -1] & 3 & 2 \\
~[0, 1, 0] &[0, 1, -1] & 3 & 3 \\
~[1, 0, 0] &[1, -1, 0] & 3 & 3 \\
~[0, 0, 1] &[0, -1, 1] & 2 & 3 \\
~[1, 0, 0] &[1, -1, 0] & 2 & 1 \\
~[1, 0, 0] &[1, -1, 0] & 1 & 2 \\
~[0, 0, 1] &[0, -1, 1] & 3 & 2 \\
~[0, 0, 0] &[0, -1, 0] & 3 & 3 \\
~[0, 0, 1] &[0, -1, 1] & 3 & 3 \\
~[0, 1, 0] &[0, 1, -1] & 2 & 3\\\hline
\end{tabular}
\end{minipage}
}

\caption{\label{tab:momenta}One choice of the initial and final three-momenta (in lattice units) and polarisations (in a cartesian basis) $\{\vec{p},\vec{p}\,',E,E'\}$ contributing to each reduced set of averaged ratios $\overline{R}$. The ordering of the rows here is the same as in Eq.~(\ref{eq:bigmats}).}
\end{table}

The system of equations determining $A_{i,0}^{(2)}(\Delta^2=1)$ for this basis, at this momentum, can be expressed as:

\pagebreak

\begin{widetext}

{\tiny
$
\begin{pmatrix}
0.604&0.0424&0&0&0&0&0.0588&0\\0.592&-2.45\times 10^{-3}&0.0785&-0.0785&6.58\times 10^{-3}&-0
.0992&-0.103&-4.15  \times 10^{-3}\\0.485&0.0429&0&0&0&0&0.0379&0\\0.481&0.0431&-3.02\times 10^{-
	5}&3.02\times 10^{-5}&-2.53\times 10^{-6}&-4.03\times 10^{-7}&0.0374&-1.69\times 10^{-8}\\0.475
&-3.29\times 10^{-3}&0.0791&-0.0791&6.59\times 10^{-3}&-0.0791&-0.0824&-3.29\times 10^{-3}\\0.3
53&-7.97\times 10^{-4}&0.0385&-0.0385&3.28\times 10^{-3}&-0.0598&-0.0631&-2.54\times 10^{-3}\\0
.347&-0.0382&0&0&0&0&0.0962&0\\0.258&0.0806&0&0&0&0&-0.0374&0\\0.258&0.0808&0&0&0&0&-0.0379&0\\
0.253&0.101&-8.60\times 10^{-4}&8.60\times 10^{-4}&-7.20\times 10^{-5}&6.32\times 10^{-7}&-0.05
88&2.65\times 10^{-8}\\0.239&-1.66\times 10^{-3}&0.0401&-0.0401&3.29\times 10^{-3}&-0.0393&-0.0
402&-1.61\times 10^{-3}\\0.238&-1.65\times 10^{-3}&0.0396&-0.0396&3.29\times 10^{-3}&-0.0396&-0
.0412&-1.65\times 10^{-3}\\0.228&-0.0581&8.30\times 10^{-4}&-8.30\times 10^{-4}&6.94\times 10^{-5}&-1.04\times 10^{-6}&0.0962&-4.33\times 10^{-8}\\0.228&-0.0379&0&0&0&0&0.0758&0\\0.0590&-0.0
109&0.139&-0.139&0.0112&-4.97\times 10^{-3}&-3.94\times 10^{-4}&-8.24\times 10^{-6}\\0.0578&-2.56\times 10^{-4}&9.42\times 10^{-3}&-9.42\times 10^{-3}&3.89\times 10^{-4}&-4.65\times 10^{-3}&
2.51\times 10^{-4}&5.25\times 10^{-6}\\0.0338&1.59\times 10^{-3}&-0.128&0.128&-0.0107&3.18\times 10^{-4}&0.0154&1.33\times 10^{-5}\\0.0183&6.36\times 10^{-3}&-1.29\times 10^{-4}&1.29\times 10^{-4}&3.84\times 10^{-4}&4.84\times 10^{-3}&5.99\times 10^{-3}&5.18\times 10^{-6}\\0.0155&-4.7
8\times 10^{-3}&-0.128&0.128&-0.0111&-4.52\times 10^{-3}&9.41\times 10^{-3}&8.14\times 10^{-6}\\1.19\times 10^{-3}&-0.0106&0.129&-0.129&0.0108&-3.22\times 10^{-4}&-6.45\times 10^{-4}&-1.35\times 10^{-5}\\0.549&2.44\times 10^{-3}&0&0&0&0&0.0895&0\\0.546&-1.88\times 10^{-3}&0.0676&-0.0676&5.69\times 10^{-3}&-0.0918&-0.0960&-3.86\times 10^{-3}\\0.498&0.0710&0&0&0&0&0.0123&0\\0.480 &-2.37\times 10^{-3}&0.0685&-0.0685&5.70\times 10^{-3}&-0.0799&-0.0828&-3.33\times 10^{-3}\\0.429&0.0714&0&0&0&0&0&0\\0.424&0.0834&-5.14\times 10^{-4}&5.14\times 10^{-4}&-4.30\times 10^{-5}&
1.33\times 10^{-7}&-0.0123&5.55\times 10^{-9}\\0.412&2.85\times 10^{-3}&0&0&0&0&0.0657&0\\0.412
&-2.85\times 10^{-3}&0.0685&-0.0685&5.70\times 10^{-3}&-0.0685&-0.0714&-2.85\times 10^{-3}\\0.4
09&-8.65\times 10^{-3}&4.61\times 10^{-4}&-4.61\times 10^{-4}&3.86\times 10^{-5}&-8.30\times 10
^{-7}&0.0771&-3.47\times 10^{-8}\\0.0674&-6.43\times 10^{-3}&0.0856&-0.0856&6.70\times 10^{-3}&
-5.55\times 10^{-3}&-8.26\times 10^{-5}&-1.73\times 10^{-6}\\0.0656&4.96\times 10^{-4}&-9.21\times 10^{-4}&9.21\times 10^{-4}&-6.37\times 10^{-6}&-0.0119&-0.0132&-5.32\times 10^{-4}\\0.0514&
-0.0685&0&0&0&0&0.0771&0\\0.0347&-0.0124&0.155&-0.155&0.0127&-3.05\times 10^{-3}&-6.00\times 10
^{-4}&-1.26\times 10^{-5}\\0.0327&5.99\times 10^{-3}&-0.0692&0.0692&-6.03\times 10^{-3}&-2.50\times 10^{-3}&5.17\times 10^{-4}&1.08\times 10^{-5}\\0.0301&4.59\times 10^{-3}&-0.0738&0.0738&-5
.95\times 10^{-3}&2.98\times 10^{-3}&0.0123&1.07\times 10^{-5}\\0.0285&-1.84\times 10^{-3}&-0.1
47&0.147&-0.0126&-2.43\times 10^{-3}&0.0143&1.24\times 10^{-5}\\0.0171&0.0685&0&0&0&0&-0.0657&0
\\0.0146&0.0920&-9.75\times 10^{-4}&9.75\times 10^{-4}&-8.17\times 10^{-5}&9.63\times 10^{-7}&-
0.0895&4.03\times 10^{-8}\\1.59\times 10^{-3}&6.43\times 10^{-3}&0.0736&-0.0736&6.61\times 10^{
	-3}&5.40\times 10^{-3}&-1.97\times 10^{-3}&-1.71\times 10^{-6}
\end{pmatrix}
$
$
\begin{pmatrix}
A^{(2)}_{1,0}(1)\\[2pt]
A^{(2)}_{2,0}(1)\\[2pt]
A^{(2)}_{3,0}(1)\\[2pt]
A^{(2)}_{4,0}(1)\\[2pt]
A^{(2)}_{5,0}(1)\\[2pt]
A^{(2)}_{6,0}(1)\\[2pt]
A^{(2)}_{7,0}(1)\\[2pt]
A^{(2)}_{8,0}(1)
\end{pmatrix}
$
$
=
\begin{pmatrix}
0.179(36) \\
0.150(38) \\
0.152(30) \\
0.154(37) \\
0.129(32) \\
0.056(31) \\
0.067(41) \\
0.056(35) \\
0.069(21) \\
0.093(36) \\
0.028(32) \\
0.041(27) \\
0.012(33) \\
0.029(30) \\
0.024(11) \\
-0.005(21) \\
-0.0056(96) \\
-0.002(11) \\
0.009(16) \\
0.0162(91) \\
0.086(26) \\
0.131(31) \\
0.155(33) \\
0.086(33) \\
0.098(16) \\
0.094(17) \\
0.088(27) \\
0.114(25) \\
0.075(27) \\
0.034(25) \\
-0.006(22) \\
-0.001(31) \\
0.022(11) \\
0.014(16) \\
0.0010(16) \\
0.0008(85) \\
0.018(23) \\
0.001(29) \\
0.005(18) \\
\end{pmatrix}
$
}
\begin{equation} {\label{eq:bigmats}} \end{equation} 

\noindent
where the numbers and uncertainties on the right hand side of the equation come from the plateau fits to averaged ratios obtained as described in the main text. The ordering of the rows is as in Table~\ref{tab:momenta}.

\end{widetext}

\section{Direct solution of form factor decomposition for electromagnetic current}
\label{app:bob}
\def\cR{{\cal R}}

Since only three form factors contribute to matrix elements of the electromagnetic current, a direct solution of the constraint equations relating ratios of three-point and two-point functions to the form factors is straightforward \cite{Hedditch:2007ex,Owen:2015gva}. This extraction is performed as a check on the more general method discussed in Section~\ref{sec:analysis}.

For each momentum transfer, ${\Delta^2}$, three ratios of two-point and three-point functions are required to extract the form factors at that momentum. At zero momentum transfer, only the $G_1$ form factor can be determined. 
In terms of the ratios
\begin{eqnarray}
\cR^i_{jk}(\vec{\Delta}) = R_{jk}(\vec{p}=\vec{\Delta},\vec{p}\,'=\vec{0},t,\tau, J^i)
\end{eqnarray}
for the currents $J^i=\bar \psi \gamma^i \psi$, where $R_{jk}(\vec{p},\vec{p}\,',t,\tau,\mathcal{O})$ is defined in Eq.~\eqref{eq:ratR} and dependence on the current and sink times is suppressed,
the generic form of the solution for the FFs can be expressed as
\begin{eqnarray}
G_X(\Delta^2) = {\cal M}_X \sum_{f=a,b,c}{ N}_{X,f} R_{X,f}.
\end{eqnarray}
Here $X=C,M,Q$ labels the Sachs form factors, which are related to the basis used in Eq.~\eqref{eq:EMFFs} by
\begin{eqnarray}
G_Q(\Delta^2) &=& G_1(\Delta^2) - G_2(\Delta^2) + (1 +\eta) G_3(Q^2), \nonumber \\
G_M(\Delta^2) &=& G_2(\Delta^2), \\
G_C(\Delta^2) &=& G_1(\Delta^2) + \frac{2}{3} \eta\, G_Q(\Delta^2) \nonumber \, .
\end{eqnarray}
One choice of the combinations $N_{X,f}$ for each momentum used, given that only zero sink momentum sequential propagators were computed, is given in Table~\ref{tab:tabby}.

\begin{table*}[!h]
	{\footnotesize
	\begin{tabular}{ccccccccc}\toprule
		$\vec{\Delta}/p$ & $X$ & ${\cal M}_X$ & $N_{X,a}$ & $R_{X,a}$ & $N_{X,b}$ & $R_{X,b}$ & $N_{X,c}$ & $R_{X,c}$ \\
		\hline
		 & C & $1/3$ & 1 & $\cR^4_{11}$ & 1 & $\cR^4_{22}$ & 1 & $\cR^4_{33}$
		\\	
		(0,0,0) & M & -- & -- & -- & -- & -- & -- & -- 
		\\	
		 & Q & -- & -- & -- & -- & -- & -- & -- 
		\\	\hline
		 & C & $\frac{2\sqrt{E m}}{3(E+m)}$ & 1 & $\cR^4_{11}$ & 2 & $\cR^4_{22}$ & -- &--
		\\	
		(1,0,0) & M & $-\frac{2\sqrt{E^3}}{p\sqrt{m}}$ & -- & -- & -- & -- & 1 & $\cR^2_{12}$
		\\	
		 & Q & $\frac{2\sqrt{E m^3}}{p^2}$ & 1 & $\cR^4_{11}$ & -1 & $\cR^4_{22}$ & -- &--
		\\	\hline
		 & C & $\frac{2\sqrt{m}}{3\sqrt{E}(E+m)}$ & $\sqrt{2} \sqrt{E^2+m^2}$ & $\cR^4_{11}$ 
		& $(2E-m)$ & $\cR^4_{33}$ & -- &--
		\\	
		(1,1,0) & M & $-\frac{2\sqrt{E(E^2+m^2)}}{p\sqrt{m}}$ & -- & -- & -- & -- & 1 & $\cR^3_{13}$
		\\	
		 & Q & $\frac{ \sqrt{m^3}}{p^2 \sqrt{E}}$ & $\sqrt{2}\sqrt{E^2+m^2}$ & $\cR^4_{11}$ & $-(E+m)$ & $\cR^4_{33}$ & -- &--
		\\	\hline
		 & C & $\frac{2\sqrt{E^2+2m^2}}{3\sqrt{Em}(E+m)}$ & $ \frac{\sqrt{3}(2E+m)}{3}$ & $\cR^4_{11}$ 
		& $-\frac{4\sqrt{3}p(E+2m)}{3(E+m)}$ & $\cR^1_{12}$ & $-\frac{4\sqrt{3}p(2E+m)}{3(E+m)}$ & $\cR^2_{12}$
		\\	
		(1,1,1) & M & $\frac{2\sqrt{Em}\sqrt{2m^2+E^2}}{\sqrt{3}p(E+m)}$ & -- & -- & 1& $\cR^1_{12}$ & -1 & $\cR^2_{12}$
		\\	
		 & Q & $\frac{2\sqrt{m}\sqrt{E^2+2m^2}}{3p^2\sqrt{3E}}$ & $(m-E)$ & $\cR^4_{11}$ 
		& $\frac{(E+2m)^2}{3p}$ & $\cR^1_{12}$ & $\frac{(E+2m)(2E+m)}{6p}$ & $\cR^2_{12}$
		\\	\hline
		 & C & $\frac{2 \sqrt{Em}}{3(E+m)}$ & $1$ & $\cR^4_{11}$ 
		& $2$ & $\cR^4_{22}$ & -- & -- 
		\\	
		(2,0,0) & M & $-\frac{\sqrt{E^3}}{p\sqrt{m}}$ & -- & -- & -- & -- & 1 & $\cR^2_{12}$
		\\	
		 & Q & $\frac{ \sqrt{Em^3}}{2 p^2}$ & $1$ & $\cR^4_{11}$ 
		& $-1$ & $\cR^4_{22}$ & -- & -- 
		\\	\hline
		 & C & $\frac{2\sqrt{ m}}{3\sqrt{E}(E+m)}$ & $\sqrt{5}\sqrt{E^2+4m^2}$ & $\cR^4_{11}$ 
		& $2(E-2m)$ & $\cR^4_{33}$ & -- &--
		\\	
		(1,2,0) & M & $-\frac{2\sqrt{E}\sqrt{E^2+4m^2}}{p\sqrt{5m}}$ & -- & -- & -- & -- & 1 & $\cR^3_{13}$
		\\	
		 & Q & $\frac{2\sqrt{m^3}}{5p^2\sqrt{E}}$ & $\sqrt{5}\sqrt{E^2+4m^2}$ & $\cR^4_{11}$ & $-(E+4m)$ & $\cR^4_{33}$ & -- &--
		\\	\hline
		 & C & $\frac{2\sqrt{2E^2+m^2}}{9\sqrt{3Em}(E+m)}$ & $(5m-2E)$ & $\cR^4_{11}$ 
		& $\frac{\sqrt{2}\sqrt{E^2+5m^2}(4E-m)}{\sqrt{2E^2+m^2}}$ & $\cR^4_{22}$ & -- &--
		\\	
		(2,1,1) & M & $\frac{\sqrt{2Em}\sqrt{m^2+2E^2}}{(E+m)(E+5m)}$ & $\frac{2\sqrt{6}}{3}$ & $\cR^4_{11}$ & $-\frac{2\sqrt{5m^2+E^2}}{\sqrt{m^2+2E^2}}$ & $\cR^4_{22}$ & $-\frac{\sqrt{6}(E+m)}{p}$ & $\cR^1_{12})$
		\\	
		 & Q & $\frac{\sqrt{m}\sqrt{2E^2+m^2}}{9\sqrt{3E}p^2}$ & $(E+5m)$ & $\cR^4_{11}$ & $-\frac{\sqrt{2}\sqrt{E^2+5m^2}(2E+m)}{\sqrt{2E^2+m^2}}$ & $\cR^4_{22}$ & -- &-- \\
		\hline\hline
	\end{tabular}
}
	\caption{\label{tab:tabby}
		One choice of ratios of two and three-point functions that allow extraction of the electromagnetic form factors. Here, $E=E(\Delta^2)$ and $p$ denotes one unit of momentum in lattice units. For each $R_{X,a}$, rotationally equivalent contributions are averaged.}
\end{table*}

\begin{figure*}
	\centering
	\subfigure[]{
		\includegraphics[width=0.48\textwidth]{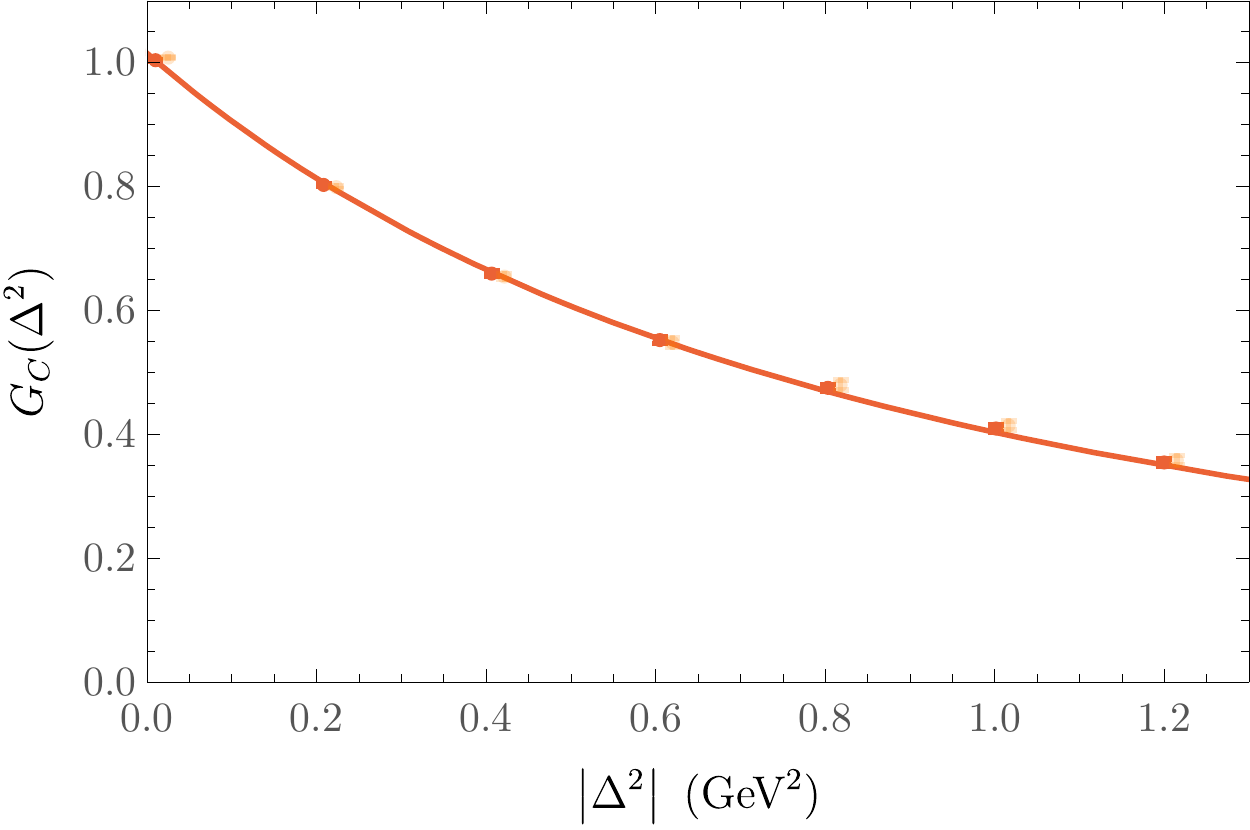}
	}
	\subfigure[]{
		\includegraphics[width=0.48\textwidth]{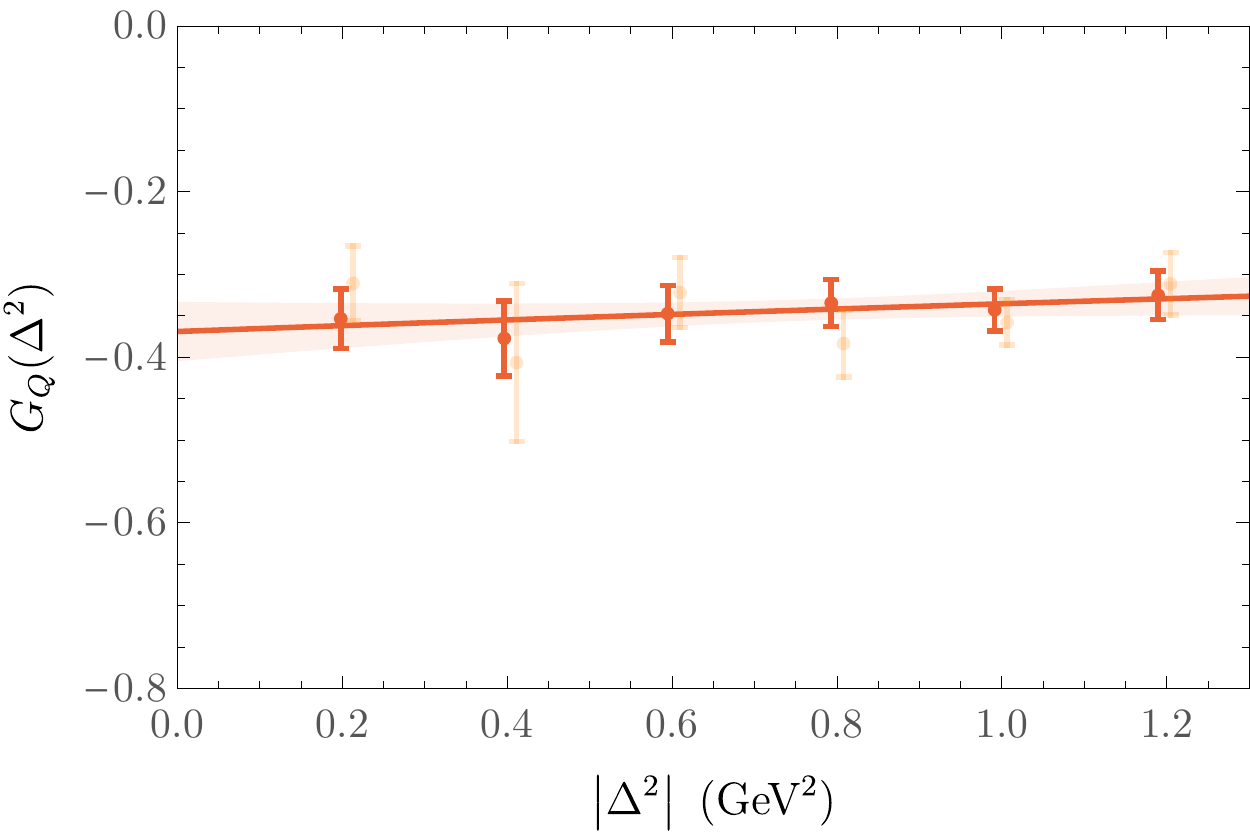}
	}
	\subfigure[]{
		\includegraphics[width=0.48\textwidth]{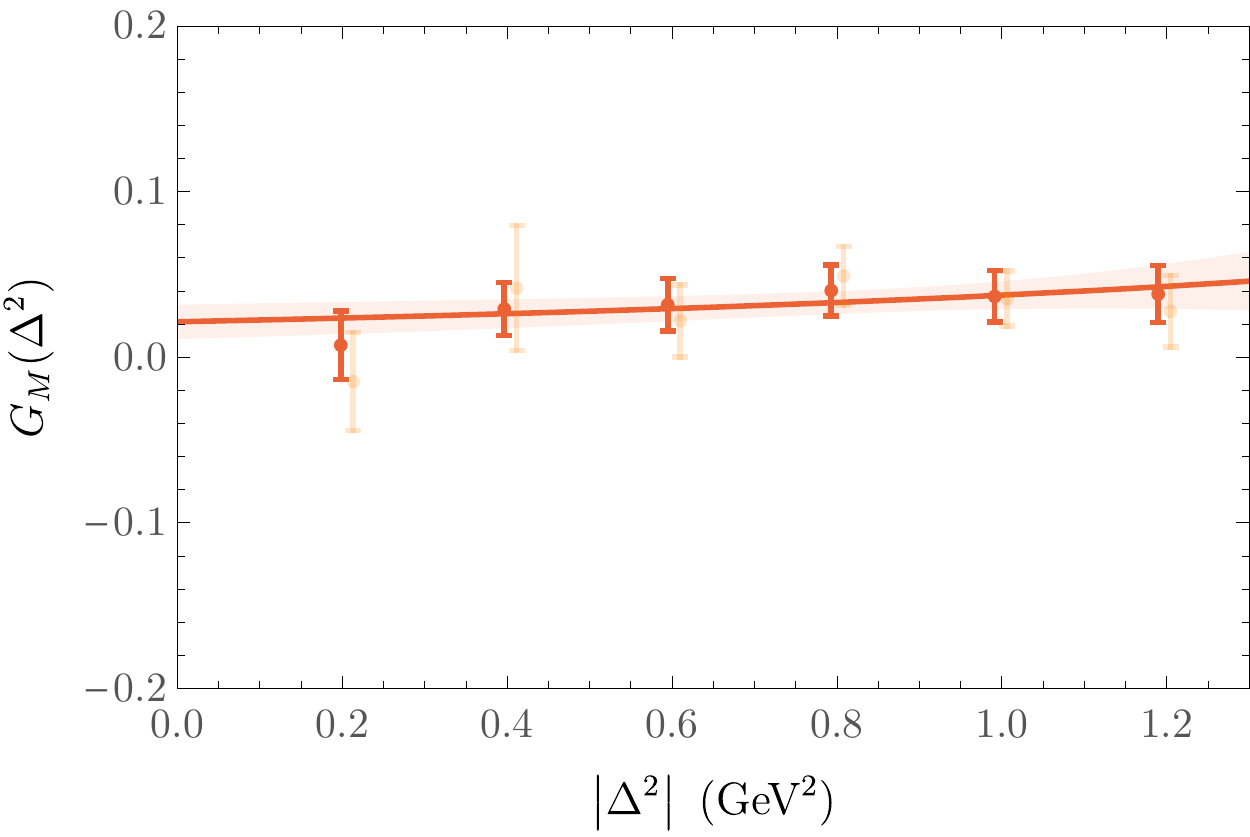}
	}
	\caption{\label{fig:QFFsCQM}Quark electromagnetic form factors determined as described in the text. As in Figs.~\ref{fig:TEGFFs} and \ref{fig:SIGFFs}, the solid red points denote the results of the full analysis, while the faded orange points (offset on the horizontal axis for clarity) show results obtained without the smoothness condition discussed at the end of Section~\ref{subsec:extract}. The bands are dipole fits to the results of the full analysis against $\left|\Delta^2\right|$.}
\end{figure*}

\bibliography{bob}

\end{document}